\documentstyle[aps,pra,twocolumn,floats,epsfig]{revtex}

\begin{document}


\draft

\wideabs{

\title{Atomic collision dynamics in optical lattices}

\author{J. Piilo,$^1$ K.-A. Suominen,$^{1,2}$ and K. Berg-S{\o}rensen$^3$}

\address{$^1$Helsinki Institute of Physics, PL 64, FIN-00014 Helsingin
yliopisto,
Finland}

\address{$^2$Department of Applied Physics, University of Turku,
FIN-20014 Turun yliopisto, Finland}

\address{$^3$Nordita and the Niels Bohr Institute,
Blegdamsvej 17, DK-2100 Copenhagen {\O}, Denmark}

\date{\today}

\maketitle

\begin{abstract}
We simulate collisions between two atoms, which move in an optical
lattice under the
dipole-dipole interaction. The model describes simultaneously the two basic
dynamical processes, namely the Sisyphus cooling of single atoms, and the
light-induced inelastic collisions between them. We consider the
$J=1/2\rightarrow
J=3/2$ laser cooling transition for Cs, Rb and Na. We find that the
hotter atoms
in a thermal sample are selectively lost or heated by the collisions,
which modifies
the steady state distribution of atomic velocities, reminiscent
of the evaporative cooling process.
\end{abstract}


}

\narrowtext

\section{Introduction}

Laser cooling and trapping techniques have made it possible to study and
manipulate samples of cold neutral atoms \cite{vanderStraten99}. It 
has lead to the
precision control of
atomic matter, and also opened new possibilities for investigations
of interaction
dynamics between atoms, especially those mediated by light \cite{Weiner99}.
  By controlling the
internal states of the atoms we obtain access to their center-of-mass motion
\cite{vanderStraten99,Castin91}.
Sisyphus cooling and polarization gradient cooling methods 
\cite{Dalibard89}, which
allow one to break
the Doppler limit for low temperatures, are based on the way polarized light
connects to the atomic states. By combining more than one laser beam we can
introduce a spatially changing polarization state into the total light field
interacting with atoms. As light introduces Stark shifts on atomic states, the
spatially changing polarization appears as a spatially changing
potential for atoms
with a suitable angular momentum state structure. In addition to the
cooling effect,
this has made it possible to build periodic or quasiperiodic
lattices \cite{Jessen96}, where the
light field traps the atoms. We describe the Sisyphus cooling and
lattice structure
in Sec.~\ref{Sisyphus}.

The basic laser cooling method is Doppler cooling \cite{vanderStraten99},
which is produced
by the random
scattering of photons absorbed from the laser beam. This cooling
mechanism is not
related to the polarization states of light. For alkali atoms this method
has a limiting temperature, the Doppler temperature $T_D=\hbar \Gamma / 2 k_B$,
where $\Gamma$
is the atomic linewidth. Using the polarization states, i.e., Sisyphus
cooling and polarization gradient cooling, one can go below $T_D$ until the
photon recoil limit, $T_R$, is reached, and creating a lattice as a
byproduct. The values of $T_R$ and $T_D$ for used elements are given in
Table~\ref{tab:AtomProperties}. Typically one reaches a thermal
equilibrium where the atoms are more or less localized at lattice
sites, but can
also move between them. The efficiency and the degree of localization have been
studied thoroughly in the past \cite{Jessen96}. Since the best 
filling ratios (number
of atoms per
site) with small-detuning lattices are on the order of 10 \% \cite{Jessen96},
one can consider the
gas sample as noninteracting. Larger filling ratios have been
achieved lately by
special techniques in far-detuned optical lattices \cite{Jaksch98}. 
Based on the
experience in
standard magneto--optical traps (MOTs), the increasing density for 
small-detuning lattices 
is expected to lead to strong loss and
heating of atoms due to collisions, which become strongly inelastic in the
presence of the near-resonant cooling light \cite{Weiner99,Holland94}.
By a ``small-detuning lattice'' we mean one that is detuned
a few atomic linewidths below the transition frequency.  

The light-assisted collisions are based on the fact that the two
slowly approaching
atoms form a quasimolecule, which the light can excite resonantly during the
approach, even though the cooling beams are clearly off the resonance 
with the atomic
transition energy which corresponds only asymptotically to the molecular state
transition energy. The resonance occurs at relatively long
distances \cite{Julienne91}, where the
dominant contribution to the molecular behavior, i.e., the
interaction potential
between atoms, comes from the dipole-dipole interaction (DDI). We
describe these
collisions in Sec.~\ref{sec:BinaryI}, and derive an expression for DDI in
Sec.~\ref{sec:DDI}. We have presented the first results of our work
in a previous
short publication~\cite{Piilo01}, where details of the derivation
were omitted, so
here we present them in full. It should be noted that one can develop
e.g. a mean
field approach to atom-atom processes via the DDI
\cite{Goldstein96,Boisseau96,Guzman98,Menotti99}. In our
approach, albeit with some limitations, we allow
the atoms to move. Furthermore, we  consider the
problem in the atom-atom basis, 
 with the full Zeeman substate structure, in the
presence of the
cooling/lattice-building laser beams with spatially changing polarization
structure. Thus our model treats the Sisyphus cooling and localization of the
atoms, and the atomic collisions dynamics consistently, within the
same framework.

In order to describe two multistate atoms moving quantum mechanically as wave
packets in position and momentum space, and being coupled both to the spatially
changing laser field as well as the vacuum field producing
spontaneous emission, we
would in principle have to use the density matrix description. This is not
computationally possible currently, but we can go around the problem
by describing
the spontaneous emission with quantum jumps. As described in
Sec.~\ref{sec:MCWP},
we use the Monte Carlo wave function (MCWF) method to build a statistical
ensemble of time
evolution histories, which approximates adequately the actual density
matrix \cite{Dalibard92,Plenio98,Molmer96}. To
implement this approach on our description of atomic collisions in
lattices is not
straightforward, and in Sec.~\ref{sec:Numerics} we describe the
details related to
numerical simulations. Inelastic collisions in MOTs have
been modelled
extensively with semiclassical models \cite{Suominen98}.
We describe these models briefly in
Sec.~\ref{sec:SC}. They provide a tool for understanding some of the
physics behind
the numerical data, and for estimating the processes affecting the
multitude of boundaries for the numerical methods.

Our Monte Carlo simulation results, presented in Sec.~\ref{sec:Results},
indicate
that the hotter atoms in our thermal sample, due to their stronger
mobility between
the lattice sites, are more likely to collide inelastically than
the colder ones.
On the other hand, the simulations, supported by semiclassical
estimates, show that
inelasticity plays a relevant role only if the atoms end up in the
same lattice site
simultaneously. In other words, the effect of the dipole-dipole
interaction remains
small if the atoms do not share the same lattice site. 
 This, of course, also
depends on the chosen laser field parameters such as intensity and
detuning.  When
the close, same-site encouter occurs, however, it is most likely a strongly
inelastic one leading to the loss of atoms. Basically, we see a
process similar to
evaporative cooling, where the hotter atoms are selectively heated or
ejected from
the lattice. It should be noted, though, that despite the fully
quantum mechanical
nature of our approach, it still omits many other effects affecting the atomic
cloud in the lattice. Photons scattered incoherently by atoms can be
reabsorbed,
which produces a radiation pressure; this process also heats the
atomic cloud as
its density and thus optical thickness increases
\cite{vanderStraten99}.

The observations made in this paper follow those from our previous
study~\cite{Piilo01}. The results given in this paper, however, have
been obtained
with an improved approach compared to Ref.~\cite{Piilo01}, and we have also
extended our studies to all the basic alkali species. Furthermore, here we
give the detailed description of our approach and its computational
aspects. Finally, the
discussion in Sec.~\ref{sec:Discussion} concludes our presentation.

\section{Sisyphus cooling and optical lattices}\label{Sisyphus}

In this Section we present the atom--laser system under study and
describe briefly
the basics of Sisyphus laser cooling of neutral atoms in an optical lattice.
Detailed review of the subject can be found in
Refs.~\cite{vanderStraten99,Dalibard89,Jessen96}.

\subsection{Atom--laser system}

\begin{figure}[tb]
    \centerline{\psfig{figure=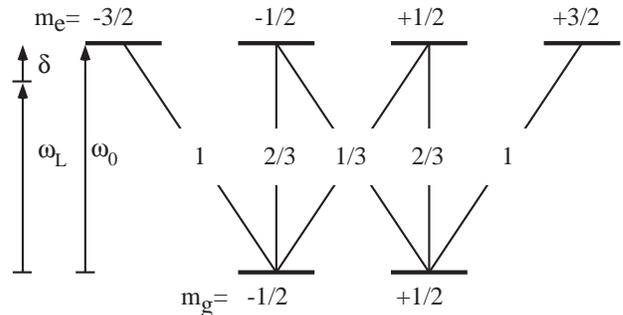,scale=0.4}}
    \caption{The level structure of a single atom. 
       We show the squares of the Clebsch-Gordan
       coefficients of corresponding transitions describing the strengths of couplings
       between the Zeeman sublevels. The difference between the laser 
       frequency $\omega_{L}$ and the atomic
       frequency $\omega_{0}$, i.e., the laser detuning, is $\delta$. }
    \label{fig:AtomicLevels}
\end{figure}

We consider here atoms having ground state angular momentum
$J_{g}=1/2$ and excited
state angular momentum $J_{e}=3/2$ corresponding to alkali metal elements when
the hyperfine structure is neglected. The resonance frequency between
the states is
$\omega_{0}$ so that $\hbar \omega_{0}=E_e-E_g$ where $E_e$ and $E_g$
are energies
of the ground and the excited states in zero field. A single atom has
two ground
state sublevels $|g_{\pm 1/2}>$ and four excited state sublevels
$|e_{\pm 3/2}>$
and $|e_{\pm 1/2}>$ where  the half--integer subscripts indicate the
quantum number
$m$ of the angular momentum along the $z$  direction, see
Fig.~\ref{fig:AtomicLevels}. The values of atomic masses that are used in our
simulations are those of cesium ($^{133}$Cs), rubidium ($^{85}$Rb) and sodium
($^{23}$Na) which are the conventional alkali metal elements used for
laser cooling
of neutral atoms, see Table~\ref{tab:AtomProperties}.

The laser field consists of two counter--propagating beams with
orthogonal linear
polarizations and with frequency $\omega_L$. The total field has a polarization
gradient in one dimension and reads 
\begin{equation}
      {\bf E}(z,t)={\cal E}_0 ({\bf e}_x e^{ik_rz} - i {\bf e}_y
      e^{-ik_rz})e^{-i\omega_L t} + c.c.,
      \label{eq:Efield}
\end{equation}
where ${\cal E}_0$ is the amplitude and $k_r$ the wavenumber. With
this field, the
polarization changes from circular $\sigma^-$ to linear and back to
circular in the
opposite direction $\sigma^+$  when $z$ changes by $\lambda_L/4$
where $\lambda_L$
is the wavelength of the lasers.

\begin{figure}[tb]
\centering
\psfig{figure=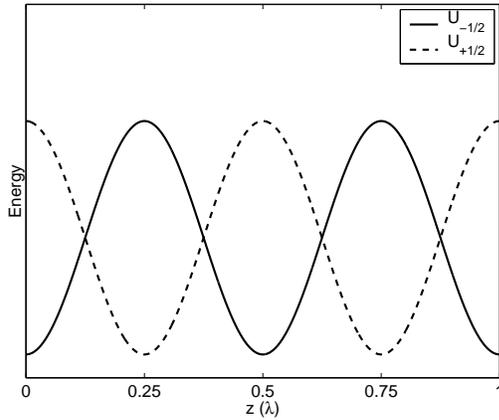,scale=0.4}
\caption[f2]{\label{fig:Lattice}
Schematic view of the optical potentials for the two ground
state Zeeman sublevels. Lattice structure is
created due to the periodic polarization gradient of the laser field.}
\end{figure}

The periodic polarization gradient of the laser field is reflected in
the periodic light shifts, AC--Stark shifts, 
of the atomic sublevels creating the optical
lattice structure. The relative strengths of the couplings between a
single ground
state sublevel and various excited state sublevels vary spatially
according to the
polarization of the light field due to unequal values of the Clebsch--Gordan
coefficients for different transitions. This induces light shifts and
produces a
periodic optical potential structure such that the shape of the light-induced
potentials is the same for the two ground state sublevels but the
potentials are
shifted spatially with respect to each other by $\lambda_L/4$, see
Fig.~\ref{fig:Lattice}.  The top of the optical potential for one sublevel
coincide with the bottom of the other one.

\subsection{Sisyphus cooling}

\begin{figure}[tb]
\centering
\psfig{figure=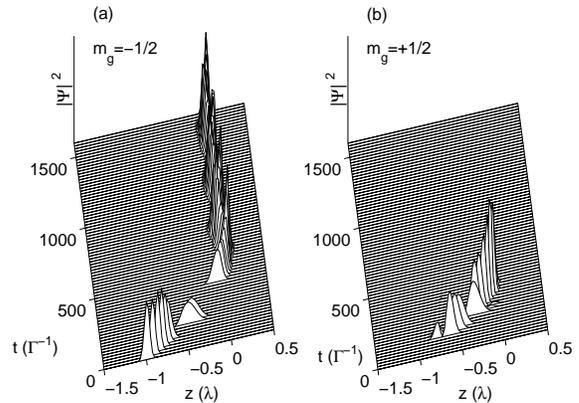,scale=0.4}
\caption[f3]{\label{fig:Localization}
Sisyphus cooling and the localization of an atom into the optical 
lattice. We show a
possible time evolution for a single atom wave packet for two ground state
Zeeman levels, (a) $m_g=-1/2$, (b) $m_g=+1/2$. The result shows the optical pumping cycles and the 
localization of a
single atom into the optical lattice.  This example forms one member of a Monte
Carlo simulation, and the discontinuous changes between the two 
ground states are
due to quantum jump events from the excited state (not shown), 
selected to happen
randomly with an appropriately weighted probability. If the run is 
repeated, the
jumps would appear at different times again.}
\end{figure}

When the atomic motion occurs in a suitable velocity range, optical
pumping of the
atom between ground state sublevels reduces the kinetic energy of the
atom~\cite{Dalibard89}.  This occurs because within the suitable
velocity range,
quantum jumps and optical pumping to another ground state sublevel
tend to occur
when the atom is near  the top of the optical potential and is transferred
to the bottom of the other one due to a quantum jump. Thus the
subsequently emitted
photon has a larger energy than the absorbed one and the kinetic
energy of the atom
is therefore reduced, and the atom is cooled. After several such
cooling cycles the
atom localizes into the optical potential well, i.e., into an optical
lattice site.
Figure~\ref{fig:Localization} shows the optical pumping cycles
between the ground
state sublevels cooling an atom, and the oscillations of the atomic wave packet
after localization into an optical lattice site.

The intensity of the laser field and the strength of the coupling
between the field
and the atom is described by  the Rabi frequency $\Omega = 2 d {\cal
E}_0 / \hbar$
where $d$ is  the atomic dipole moment of the strongest
transition between the ground and excited states. The detuning of the
laser field
from the atomic  resonance is given by $\delta = \omega_L - \omega _{0}$.
As a unit for $\Omega$ and $\delta$ we use the atomic
linewidth $\Gamma$.

\begin{table}
\caption[t1]{\label{tab:AtomProperties}
Atomic properties. Masses $M$ in a.u. and linewidth energies $\epsilon = (\hbar
\Gamma) /  E_{r}$
(for the definition of the recoil unit $E_r$ see Table~\ref{tab:Units}).
The Doppler temperature $T_D$ and the recoil temperature
$T_R=\left( \hbar^2 k_r^2 \right) / M k_B$
in $\mu K$. Here $k_r$ is the wavenumber of the laser.}
\begin{tabular}{llllll}
     property    & Cs   & Rb   & Na  \\
     \hline
     $M$           & 133  & 85   & 23  \\
     $\epsilon $ & 2400 & 1600 & 400 \\
     $T_D$       & 120  & 142  & 238 \\
     $T_R$       & 0.20 & 0.37 & 2.4 \\
\end{tabular}
\end{table}

\subsection{Localization in lattice}

When the steady state is reached after a certain period of cooling,
atoms are to a
large extent localized into the lattice sites. In this study we deal with
near--resonant bright optical lattices where the laser field is detuned a few
atomic linewidths to the red of the atomic transition. The laser parameters
$\Omega$ and $\delta$ determine if the lattice is in the
``jumping'' or in the
``oscillating'' regime, depending on the average  number of atomic
oscillations in a
single lattice site before the atom is optically pumped to neighboring
sites~\cite{Castin90}. It must be noted that tight localization and
occupation of
the lowest vibrational levels of a periodic lattice potential
increases the optical pumping
time $\tau_p$ and the time of localization within a single lattice site becomes
longer compared to the semiclassical values presented in
Table~\ref{tab:Parameters}.

We are interested in the effect of inelastic collisions between atoms in the
presence of near--resonant light~\cite{Weiner99} in optical lattices. These
collisions occur when two atoms occupy the same lattice site. To observe
efficiently the effect of inelastic  collisions we have chosen $\Omega$ and
$\delta$ in most of the simulations so that the lattice is in the ``jumping''
regime, i.e., semiclassically speaking the atoms on average do not
have time for a
single full oscillation before  optical pumping transfers them to a neighboring
lattice site.

Steady state properties of the atomic cloud in the lattice are characterized
e.g.~by the average kinetic energy per atom, the spatial probability
distribution
or the momentum probability distribution. These are results which we
obtain from
our simulations. We keep the detuning fixed $(\delta=-3 \Gamma)$ and
vary the Rabi
frequency $\Omega$, which gives various values for the optical potential
modulation depth
\begin{equation}
     U_0=-\frac{2}{3}\hbar \delta s_0, \label{eq:U_0}
\end{equation}
where $s_0$ is the saturation parameter given by
\begin{equation}
     s_0=\frac{\Omega^2/2}{\delta^2+\Gamma^2/4}. \label{eq:s_0}
\end{equation}
The spatially modulated optical potentials are
\begin{eqnarray}
     U_-=U_0\sin^2(k_rz), \nonumber \\
     U_+=U_0\cos^2(k_rz), \label{eq:U+-}
\end{eqnarray}
for ground states $m_g=-1/2$ and  $m_g=+1/2$ respectively~\cite{Castin90}.
The parameters used in our simulations along with relevant lattice properties
are summarized in Table~\ref{tab:Parameters}.

\begin{table}
\caption[t2]{\label{tab:Parameters}
Laser parameters used in the simulations and the corresponding
lattice properties:
Detuning $\delta$, Rabi frequency $\Omega$, lattice modulation depth $U_0$,
semiclassical average number of oscillations in a lattice site $N_{osc}=
\Omega_{osc} \tau_p$,  and saturation parameter $s_0$. We use the semiclassical
average oscillation frequency $\Omega_{osc}$
discussed in ~\cite{Castin90}. Units
are given in
parenthesis and the simulations are  labeled by the element and the
lattice depth. }
\begin{tabular}{llllllll}
       $\displaystyle{\Omega} (\Gamma)$
     & $\displaystyle{\delta} (\Gamma)$
     & $\displaystyle{U_0} (E_r)$
     & $\displaystyle{N_{osc}}$
     & $\displaystyle{s_0}$
     & Simulation label \\
     \hline
     1.2 & -3.0 & 374  & 0.93 & 0.08 & Cs374  \\
     1.5 & -3.0 & 584  & 0.74 & 0.12 & Cs584  \\
     2.5 & -3.0 & 1621 & 0.47 & 0.33 & Cs1621 \\
     1.8 & -3.0 & 560  & 0.76 & 0.18 & Rb560  \\
     2.8 & -3.0 & 339  & 0.98 & 0.42 & Na339  \\
     3.5 & -3.0 & 530  & 0.78 & 0.66 & Na530  \\
\end{tabular}
\end{table}

Collisions and radiative heating increase the relative velocity between the
atoms~\cite{Weiner99}. This heats up the atoms and  it is possible
for a colliding
pair to escape from the lattice.  One can calculate semiclassically
the critical
momentum $p^{sc}_{c}$ giving the point in momentum ($p$) space where
the cooling
force has its maximum value~\cite{Dalibard89}. In
Section~\ref{subsec:escape} below,
we discuss for which values of the momentum $p_{c}$ we may neglect energetic
histories and consider the corresponding atoms lost from the lattice.

\section{Binary interactions between cold atoms} \label{sec:BinaryI}

In this Section we give a simple description of collisions between
two cold atoms in
the presence of near resonant light~\cite{Weiner99} and discuss the
background of
this phenomenon to occur in an optical lattice.

\subsection{Binary interactions} \label{subsec:BinInt}

In this study, we consider atomic gases with an occupation density of 
$\rho_o = 25$~\%, i.e., every fourth lattice site is occupied and the average distance
between two atoms is $z_a=\lambda$. For Cs this corresponds to a
density of $1.62
\times 10^{12}$ atoms/cm$^3$. This atomic gas density is low enough
that collisions
can be treated as binary processes: i) The collision range is an
order of magnitude
smaller than $z_a$, and ii) for a Cs atomic mass, atoms with typical maximum
velocities produced in our simulations need to have evolved over a
time larger than
$75\,\Gamma ^{-1}$ to travel a distance $z_a$ and they scatter a large enough
number of photons that there is negligible memory  effects between
two collision
events. Thus, the binary collision picture is justified in our calculations.

\begin{figure}[tb]
\centering
\psfig{figure=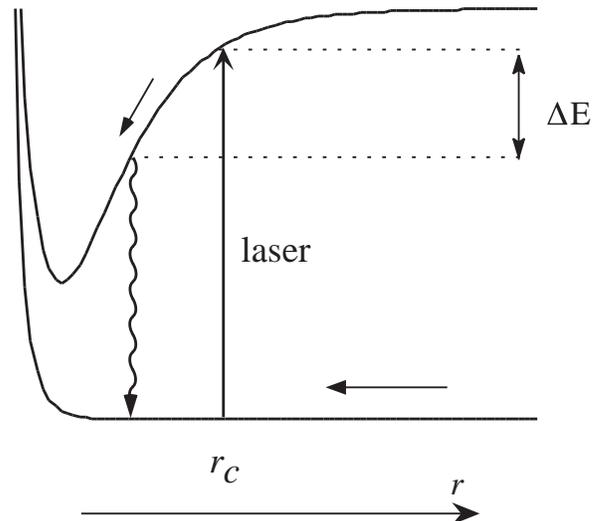,scale=0.7}
\caption[f4]{\label{fig:BinaryCollision}
Radiative heating of colliding atoms. The quasimolecule is excited at 
the Condon
point $r_c$ and accelerated on the upper level before spontaneous 
decay terminates
the process. }
\end{figure}

Let us consider two atoms with a temperature around or below the Doppler
temperature $T_{D}$. If such
two atoms collide
they form a quasimolecule which a near resonant light may excite when the atoms
approach each other. This occurs at an internuclear distance called
the Condon point
$(r_c)$ where the excited state electronic molecular potential 
becomes resonant
with the ground state potential 
 as displayed in Fig.~\ref{fig:BinaryCollision}.

We neglect the hyperfine structure of the atoms and do not consider the ground
state hyperfine structure changing collisions, but concentrate on the
effects based
on radiative  heating and escape of the colliding pair. In this process, the
resonant excitation of a quasimolecule terminates in spontaneous decay and
the colliding pair of atoms gains kinetic energy due to acceleration on an
attractive molecular excited state before decay occurs, see
Fig.~\ref{fig:BinaryCollision}. In principle it is also possible to loose atoms
from the trap via fine structure changing collisions~\cite{Weiner99}.
In this work,
the loss fraction of fine structure changing  collisions is assumed
to be negligible
compared to  the radiative escape mechanism. The small detuning of
the lasers makes
$r_c$ very large, and the probability for surviving to  the small
internuclear distances
required for the fine structure change is rather small.
Furthermore, the energy
increase caused by this mechanism is large and leads mainly to loss of atoms,
contributing to heating only via secondary collisions. These secondary
collisions are still rather rare
in the low-density gas samples of laser cooled atoms. 

An essential ingredient in the kinetic energy increasing collision
process is the
excitation of the large fraction of the population into the attractive excited
state. The excited state population fraction in turn depends on the relative
velocity between the  interacting atoms when they reach $r_c$ for the
attractive
molecular states. For the collisions in the lattice, the relative
velocity in turn
depends on the optical lattice modulation depth. The deeper the
lattice, the higher
the relative velocity of the atoms when they end up in the same
lattice site and
collide. We consider here lattices with modulation depths in the
range $339E_r\leq
U_0\leq 1621E_r$, where $E_r$ is the recoil energy, cf.\ Table~\ref{tab:Units}.
Thus, the relative velocities before a collision remain low, which keeps the
excitation probability large. The small detuning keeps the excitation
probability 
large also  since the excited state slope decreases with the detuning; this
increases the interaction time for moving atoms at the vicinity of
the resonance
point $r_c$.

\subsection{Collisions in lattices}

Theoretical and experimental collision studies in MOTs show
that the atomic cloud is heated by the radiative mechanism described
above. Atoms
may also escape from a MOT by this mechanism~\cite{Weiner99,Holland94}. Thus
these collisions
set a limit for atomic densities and temperatures of the cloud in a
MOT when the
density is increased so that binary interactions begin to have a clear effect.
Typical densities achieved in MOTs are around $10^{11}$ atoms/cm$^3$ and
temperatures around or below the Doppler cooling limit~\cite{Weiner99}.

Similar effects are expected in an optical lattice when the
occupation density of
the lattice increases. What is not directly expected
is that there is a parameter region
where a possible cooling process in a dense lattice occurs due to collisions.
This is due to the fact that the colliding pair of atoms carry more
kinetic energy
than localized ones and during a collision they almost always gain
sufficient energy
to escape from the lattice. Thus those atoms that have not collided
inelastically
and remain in a lattice, have less  kinetic energy per atom.
Moreover, they can also
thermalize via elastic ground-ground collisions. In our study, we neglect the
rescattering of photons and consequently the total pattern of cooling
and heating
is not studied. Here we only consider the effects of collisions. The complete
problem is simply not computationally tractable within our framework.

Atomic interactions in lattices are usually modelled assuming fixed positions
for both atoms and calculating how the atomic energy levels are shifted by the
interaction~\cite{Goldstein96,Boisseau96,Guzman98,Menotti99}. These
static models
ignore the dynamical nature of  the collision processes described here. When
allowing the atoms to move the problem becomes complicated and computationally
extremely tedious. To make numerically feasible calculations, we have fixed
one atom and allow the other one to move freely,  as described
further in Sec.~\ref{subsec:Fix}.

\section{ Atomic basis formulation and dipole-dipole
interaction}\label{sec:DDI}

In this section we describe the two--atom product state
basis~\cite{Cohen-Tannoudji77} and the dipole--dipole  interaction
(DDI) between
two atoms in our one-dimensional $(1D)$ study.

\subsection{Atomic basis formulation}

We do not use the adiabatic elimination of the excited states, which
is typically
employed in order to simplify the equations for atomic motion~\cite{Petsas99}.
By keeping the excited states in the calculation we are able to  account for
the dynamical nature of atomic interactions and the radiative
heating/escape mechanism.

In general the product state basis vectors are:
\begin{equation}
     |j_{1} {m_{1}}\rangle_{1}
     |j_{2}  {m_{2}}\rangle_{2}, \label{eq:Basis}
\end{equation}
where $j_{1}$ and $j_{2}$ denote the ground or excited state (in our 
case $g$ for
the ground state $^2S_{1/2}$, $e$ for the $^2$P$_{3/2}$ excited state) and
$m_{1}$, $m_{2}$
denote the quantum number for the component of $j$ along the
quantization axis $z$
for atom 1 and 2 respectively. The total number of states is $6\times 6=36$.

We have to fix the position of one atom,  as described
in Sec.~\ref{subsec:Fix}.
If the position
of atom 1 is fixed, the binary system wave function depends now only
on the position
of the moving atom 2
\begin{equation}
     |\psi(z_2,t)\rangle  = \sum_{j_{1},j_{2},m_{1},m_{2}}
     \psi^{j_{1},m_{1}}_{j_{2},m_{2}}(z_2,t) |j_{1} {m_{1}}\rangle_{1}
     |j_{2} {m_{2}}\rangle_{2}. \label{eq:Psi}
\end{equation}
The atomic spatial dimensionality of the problem is reduced from two
to one. The
relative coordinate $z$ between atoms is now $z=z_2-z_f$ where $z_f$
is the position
of the fixed atom, see Sec.~\ref{subsec:Fix}.

In the atomic product state basis~\cite{Cohen-Tannoudji77}, our
system Hamiltonian
is
\begin{equation}
     H_S=H_1+H_2+V_{dip} \label{eq:HS} .
\end{equation}
Here, $V_{dip}$ includes the interaction between the atoms and
$H_1=\widetilde{H}_1
\otimes \openone_2$ and $H_2=\openone_1 \otimes \widetilde{H}_2$, where the
operators $\openone_\alpha$ are unity operators in atom $\alpha$ subspace, and
the single atom Hamiltonian for atom $\alpha$ ($\alpha=1,2$) is,
after the Rotating Wave Approximation (RWA),
\begin{equation}
     \widetilde{H}_\alpha =  \frac{p_{\alpha}^{2}}{2M} - \hbar \delta
P_{e,\alpha}
      + \widetilde{V}_\alpha \label{eq:HAlpha}.
\end{equation}
Here, $P_{e,\alpha} =\sum_{m=-3/2}^{3/2} |e_m \rangle_{\alpha}~_{\alpha} \langle
e_m|$, and the interaction between a single atom $\alpha$ and the field is
\begin{eqnarray}
     \widetilde{V}_\alpha&=& -i\frac{\hbar\Omega}{\sqrt{2}} \sin(kz_{\alpha})
     \left\{|e_{3/2} \rangle_{\alpha}~_{\alpha}\langle g_{1/2}| \right.
\nonumber \\
     && + \frac{1}{\sqrt{3}} \left.
     |e_{1/2} \rangle_{\alpha}~_{\alpha}\langle g_{-1/2}|\right\} \nonumber \\
     && +\frac{\hbar\Omega}{\sqrt{2}}\cos(kz_{\alpha})
     \left\{|e_{-3/2} \rangle_{\alpha}~_{\alpha}\langle g_{-1/2}|
\right. \nonumber \\
     &&  + \frac{1}{\sqrt{3}} \left.
     |e_{-1/2} \rangle_{\alpha}~_{\alpha}\langle g_{1/2}|\right\} +h.c.,
     \label{eq:VAtomLaser}
\end{eqnarray}
where $z_\alpha$ is the position operator of atom $\alpha$.

\subsection{Resonant dipole--dipole interaction}

In order to get the DDI potential, $V_{dip}$, in Eq.~(\ref{eq:HS}), we have
calculated the Master Equation for the atom and laser field in question, and
identified the DDI.  Our approach follows the lines of Appendix A in
Ref.~\cite{Lenz93}.  As it is beyond the scope of this paper to go through the
derivation of the DDI potential in detail, we shall refer to equation (Ax) in
Ref.~\cite{Lenz93} as Eq.~(LMAx). We identify $V_{dip}$ as the terms similar to
$\Delta_{11}$ and $\Delta_{22}$ with 
$\left< n_\omega + 1 \right> = 1$ in Eq.~(LMA21).

First, it is convenient to write the non-interacting system
Hamiltonian $H_1+H_2$ in
a basis of center-of-mass and relative coordinates:
\begin{equation}
     {\bf P}={\bf p}_1+{\bf p}_2, \;\;\;
     {\bf p}=\frac{{\bf p}_2-{\bf p}_1}{2}. \label{eq:coordinates}
\end{equation}
With these coordinates, the interaction potential with the laser field,
$V=\widetilde{V}_1 \otimes \openone_2 + \openone_1 \otimes \widetilde{V}_2$,
reads
\begin{eqnarray}
     V&=&-i\frac{\hbar\Omega}{\sqrt{2}}\sin (kZ) \cos \left(k\frac{z}{2}\right)
     \left(S_{+,+}^{1}\otimes\openone_2+
     \openone_1\otimes S_{+,+}^{2}\right)\nonumber \\
     &&+i\frac{\hbar\Omega}{\sqrt{2}}\cos (kZ) \sin \left(k\frac{z}{2}\right)
     \left(S_{+,+}^{1}\otimes\openone_2
     -\openone_1\otimes S_{+,+}^{2}\right)  \nonumber \\
     &&+i\frac{\hbar\Omega}{\sqrt{2}}\cos (kZ) \cos \left(k\frac{z}{2}\right)
     \left(S_{+,-}^{1}\otimes\openone_2+\openone_1\otimes S_{+,-}^{2}\right)
     \nonumber \\
     &&+i\frac{\hbar\Omega}{\sqrt{2}}\sin (kZ) \sin \left(k\frac{z}{2}\right)
     \left(S_{+,-}^{1}\otimes\openone_2-\openone_1\otimes
S_{+,-}^{2}\right) \nonumber
     \\ &&  + h.c., \label{eq:Vcoordinates}
\end{eqnarray}
where $Z$ and $z$ are the center-of-mass and relative coordinates
along the $z$-axis
and
\begin{equation}
      S_{+,q}^{\alpha}=\sum_{m=-1/2}^{m=1/2} CG_{m}^{q}
      |e_{m+q}\rangle_\alpha ~_\alpha\langle g_m|.
      \label{eq:S+q-alpha}
\end{equation}
Here $CG_{m}^{q}$ are the appropriate Clebsch-Gordan coefficients and
$q$ is the
polarization label in the spherical basis. We rewrite similarly the
interaction with
the vacuum electromagnetic field in terms of the relative
coordinates~(\ref{eq:coordinates}). The DDI terms are identified after we have
considered the damping part of $\dot{\rho}$ [cf.~Eq.~(LMA17)]  in the
derivation of
the Master Equation for our two-atom system.

Following~\cite{Lenz93}, we note that the DDI potential is found as
\begin{eqnarray}
     V_{dip} &=&- \frac{3}{8} \hbar \Gamma \frac{1}{\pi}
     \int_{0}^{\infty} d \omega \left( \frac{\omega}{\omega_0} \right)^3
     P\left(\frac{1}{\omega-\omega_0} \right) \times \nonumber \\
     && \left \{ j_0 \left( \omega \frac{r}{c} \right)
     \left( \frac{1}{3} ({\cal S}_{++}{\cal S}_{-+} +
     {\cal S}_{+-} {\cal S}_{--} ) - \frac{2}{3} {\cal S}_{+0}
     {\cal S}_{-0} \right) \right. \nonumber \\
     && + j_2 \left( \omega \frac{r}{c} \right)
     \left[ P_2( \cos \theta_r ) \left( -\frac{2}{3} (
     {\cal S}_{++} {\cal S}_{-+} + {\cal S}_{+-}{\cal S}_{--}) \right. \right.
     \nonumber\\
     &&\left. + \frac{4}{3} {\cal S}_{+0}{\cal S}_{-0} \right)
     +\frac{1}{3\sqrt{2}} P_{2}^{1} (\cos \theta_r) \cos \phi_r \times
     \label{eq:Vdipintegral} \\
     && \left(-{\cal S}_{++} {\cal S}_{-0} + {\cal S}_{+0}{\cal S}_{--}  -{\cal
     S}_{+0}{\cal S}_{-+}+{\cal S}_{+-}{\cal S}_{-0}\right) \nonumber \\
     && +\frac{1}{3} P_{2}^{2} (\cos \theta_r) \cos 2\phi_r
     \left. \left. \left({\cal S}_{++} {\cal S}_{--}+{\cal S}_{+-}{\cal S}_{-+}
     \right) \frac{}{}\!\right] \right\},\nonumber
\end{eqnarray}
where $P(x)$ is Cauchy's principal value, $j_{\ell}$ are spherical
Bessel functions
of the first kind, $P_{2}$ is Legendre polynomial, and $P_{m}^{n}$
are associated
Legendre functions. The angles $\theta_r$ and $\phi_r$ are the angles of the
relative coordinate ${\bf r}$ in the spherical basis. We have also
introduced the
operators
\begin{equation}
      {\cal S}_{+q} {\cal S}_{-q'} \equiv
      \left( S_{+,q}^{1}S_{-,q'}^{2} + S_{+,q}^{2} S_{-,q'}^{1}
      \right). \label{eq:S+qS-q}
\end{equation}
where $S_{-,q}^{\alpha}=\left( S_{+,q}^{\alpha}\right)^{\dagger}$.

Thus, we need to calculate integrals of the type
\begin{equation}
     {\cal I}_{\ell} (q_0 r) = \frac{1}{\pi} \int_{0}^{\infty} d\omega
     \left( \frac{\omega}{\omega_0} \right)^3
     P\left( \frac{1}{\omega-\omega_0} \right)
     j_{\ell} \left( \omega \frac{r}{c} \right),
     \label{eq:Pvalueintegrals}
\end{equation}
with $q_0 = \omega_0/c$. We may change
the lower limit
in the integral to $-\infty$, enabling us to calculate the integral by contour
integration~\cite{Berman97}. The results are
\begin{eqnarray}
     {\cal I}_0 (q_0 r) &=& \frac{\cos q_0 r}{q_0 r}  \label{eq:Pvintegrals}\\
     {\cal I}_2 (q_0 r)&=& -\frac{\cos q_0 r}{q_0 r} +
     3 \left( \frac{\sin q_0 r}{(q_0 r)^2} + \frac{\cos q_0 r}{(q_0 r)^3}
     \right), \nonumber
\end{eqnarray}
and the three-dimensional DDI potential is
\begin{eqnarray}
     V_{dip} &=&- \frac{3}{8} \hbar \Gamma \left\{
     \frac{1}{3} \frac{\cos q_0 r}{q_0 r}
     (1-2P_2 (\cos \theta_r))\times \right. \nonumber \\
     && \left({\cal S}_{++}{\cal S}_{-+} +
     {\cal S}_{+-} {\cal S}_{--}  - 2 {\cal S}_{+0}{\cal S}_{-0} \right)
     \nonumber \\
     && - 2\left(\frac{\sin q_0 r}{(q_0 r)^2} + \frac{\cos q_0 r}{(q_0 r)^3}
     \right) P_2(\cos \theta_r) \times \nonumber \\
     && \left( {\cal S}_{++}{\cal S}_{-+} +
     {\cal S}_{+-} {\cal S}_{--}  - 2 {\cal S}_{+0}{\cal S}_{-0} \right)
     \nonumber \\
     && + \frac{1}{3} \left(-\frac{\cos q_0 r}{q_0 r}
     +3(\frac{\sin q_0 r}{(q_0 r)^2} + \frac{\cos q_0 r}{(q_0
r)^3})\right) \times
     \nonumber \\
     && \left[ \frac{1}{\sqrt{2}} P_{2}^{1} (\cos \theta_r) \cos \phi_r
\times \right.
     \nonumber \\ && \left(-{\cal S}_{++} {\cal S}_{-0} + {\cal
S}_{+0}{\cal S}_{--}
     -{\cal S}_{+0}{\cal S}_{-+}+{\cal S}_{+-}{\cal S}_{-0}\right) \nonumber \\
     && +  P_{2}^{2} (\cos \theta_r) \cos 2\phi_r \times \nonumber \\
     && \left. \left. \left({\cal S}_{++} {\cal S}_{--}+{\cal
S}_{+-}{\cal S}_{-+}
     \right)\frac{}{}\! \right] \right\}.
\end{eqnarray}

\begin{figure}[tb]
\centering
\psfig{figure=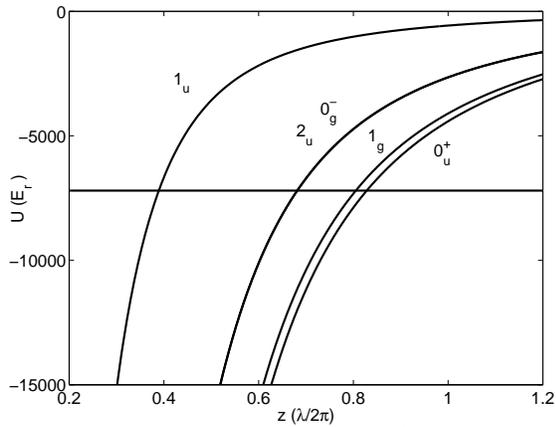,scale=0.4}
\caption[f5]{\label{fig:MolPots}
The shifted ground state and the attractive excited state [labeled by
Hund's case (c) notation] molecular potentials of Cs for $\delta=-3.0 \Gamma$.}
\end{figure}

If the two atoms are positioned on the $z$-axis, the DDI potential
reduces to the
one-dimensional potential
\begin{eqnarray}
      V_{dip}^{axis}&=&\frac{3}{8}\hbar\Gamma \left\{ \frac{1}{3}
      \frac{\cos q_0 r}{q_0 r}
      +2\left[ \frac{\sin q_0 r}{(q_0 r)^2} + \frac{\cos q_0 r}{(q_0 r)^3}
      \right] \right\} \times \nonumber \\
      &&~~~\left( {\cal S}_{++}{\cal S}_{-+} + {\cal S}_{+-}{\cal S}_{--}
      - 2{\cal S}_{+0}{\cal S}_{-0} \right).
      \label{eq:VDipAxis}
\end{eqnarray}
By diagonalizing $V_{dip}$ it is possible to obtain the molecular potentials
shown in Fig.\ \ref{fig:MolPots}. One also notes that the DDI induces the
$\pi$--polarization couplings which the laser fields do not do here.

\section{Monte Carlo wave function method}\label{sec:MCWP}

In this section we describe briefly the main features of the Monte
Carlo wave function
(MCWF) method~\cite{Dalibard92} which was developed for problems in
quantum optics
and discuss the implementation of the method to solve the cold
collision problem
in optical lattices.

\subsection{Basic Monte Carlo method}

Various types of Monte Carlo (MC)
methods~\cite{Dalibard92,Plenio98,Molmer96} have
been developed for problems where a direct analytical or numerical quantum
mechanical solution of the density matrix Master Equation is very difficult or
impossible due to the complexity of the problem.  Complexity usually
arises because
of the coupling of the system studied to a reservoir with a large
number of degrees
of freedom  and also because of a large number of elements in the
system density
matrix. Problems of this kind are common in laser cooling of neutral atoms.
Various types of quantum approaches are possible in 2D systems \cite{Castin94}
but in 3D a full quantum treatment of  laser cooling of atoms has 
only been given
in terms of the Monte Carlo method~\cite{Molmer95}.

The core idea of MCWF method is the generation of a large number of single wave
function histories including stochastic quantum jumps of the system studied.
Solutions for the steady state density matrix and system properties can then be
calculated as ensemble averages of single histories.

To generate single histories of the system wave function
$|\psi\rangle$, one solves
the time-dependent Schr\"odinger equation
\begin{equation}
     i \hbar \frac{\partial|\psi\rangle}{\partial t}=
     H|\psi\rangle. \label{eq:Schrodinger}
\end{equation}
Here the non--Hermitian Hamiltonian $H$ is
\begin{equation}
     H=H_{S}+H_{DEC} \label{eq:H}
\end{equation}
where $H_{S}$ is the system Hamiltonian,  Eq.~(\ref{eq:HS}) in our
case,  and the
non-Hermitian part $H_{DEC}$ includes the decay part. $H_{DEC}$ is
constructed from
appropriate jump operators $C_j$ corresponding to a decay channel $j$
and to the
detection scheme of the system. The general form of the non-Hermitian
part reads
\begin{equation}
     H_{DEC}=-\frac{i\hbar}{2}\sum_{j}C_j^{\dagger}C_j. \label{eq:HDec}
\end{equation}

During a time evolution step $\delta t$ the norm of the wave function
may shrink due
to $H_{DEC}$ and the amount of shrinking gives the probability of a
quantum jump to
occur during the short interval $\delta t$. Based on a random number one then
decides whether a quantum jump occurred or not. Before the next time
step is taken,
the  wave function of the system is renormalized. In the case that a
jump occurs,
one performs a rearrangement of the wave function components
according to the jump
operator $C_j$, corresponding to decay channel $j$, before renormalization of
$|\psi\rangle$.

\subsection{Time evolution}

A natural termination for a simulation occurs if a steady state
appears. One must
note that the time to reach steady state varies even for the same
system studied
when the laser parameters are changed. Therefore one has to be
careful to have long
evolution times to make sure that the steady state is reached and ensemble
averaging can be done in a reliable way.

We solve the time--dependent Schr\"odinger equation Eq.~(\ref{eq:Schrodinger})
by the split operator--Fourier transform method~\cite{Garraway95}.
Formally solving
Eq.~(\ref{eq:Schrodinger}) over $\delta t$ gives
\begin{equation}
     |\psi(t_0+\delta t)\rangle = U |\psi(t_0)\rangle \label{eq:Psitdt}
\end{equation}
where the time evolution operator U reads
\begin{equation}
     U=\exp\left(-\frac{i H\delta t}{\hbar}\right) \label{eq:Uformal}.
\end{equation}

We split the time evolution operator $U$ including the Hamiltonian $H$  of 
Eq.~(\ref{eq:H})  into three parts as $H=H_{V}+H_{K}+H_{D}$. When $H$ is in
matrix form, $H_{V}$ has an off-diagonal part accounting for the atom--field
coupling and the interaction between atoms, $H_{K}$ is the diagonal
kinetic part and
$H_{D}$ includes the non-kinetic diagonal part, i.e., decay and detuning.

For non-commuting operators $A$ and $B$ we can write to second order
accuracy~\cite{Garraway95}
\begin{equation}
     \exp\left(A+B\right) \simeq
     \exp\left(A/2\right)\exp\left(B\right)\exp(A/2) \label{eq:ExpA+B}.
\end{equation}
As we take many consecutive time steps during the evolution, we
finally approximate
the wave function at time $t_0 + n \delta t$ by
\begin{equation}
     |\psi(z,t_0+n\delta t)\rangle
     \simeq \left[\prod_{k=0}^{n-1}U_{V}U_{D}^{1/2}U_{K}U_{D}^{1/2}\right]
     |\psi(z,t_0)\rangle  \label{eq:U}.
\end{equation}
Here, $U_{D}\!=\!\exp(-iH_{D}\delta t/\hbar)$ and $U_{K}\! = \!{\cal
F}^{-1} \exp
(-i\delta t \frac{\hbar k^2}{2M})$ where ${\cal F}$ and ${\cal
F}^{-1}$ denote the
Fourier and inverse Fourier transforms. Finally, $U_{V}$ can be written as
$U_{V}=S\exp(D)S^{-1}$ where $S$ contains eigenvectors and $D$ eigenvalues of
$H_{V}$. $U_V$ corresponds now to a change of basis,  multiplication by
exponentials of eigenvalues and change of basis back to the product
state basis.
The above form for  the temporal evolution of $|\psi \rangle$ is
straightforward to
implement and fast on a computer,  e.g., 20\% faster than the 
Crank--Nicholson method. 

\subsection{Decay channels}

A single atom has six different ways to spontaneously emit a photon
so the total
number of decay channels is 12 for two atoms (channels $1-6$ for atom 1, $7-12$
for atom 2). For each decay channel, $j$, the jump probability is given by
\begin{equation}
     P_j=\delta t \langle\psi|C_j^{\dagger}C_j|\psi\rangle, \label{eq:Jp}
\end{equation}
where the jump operators $C_j$ are constructed from single particle jump
operators.   In the single particle subspace for atom $\alpha$
and decay channel $j$ we have
\begin{equation}
     \widetilde{C}_j^{\alpha}=CG_j\sqrt{\Gamma}~|g_{\alpha} m_{\alpha}
     \rangle_{\alpha}~_{\alpha}\langle e_{\alpha} m_{\alpha}|, \label{eq:Cj}
\end{equation}
where $e_{\alpha} m_{\alpha}$ labels the excited level from which, 
and $g_{\alpha} m_{\alpha}$
the ground level
to which jump occurs. Extension to the product state basis is
simple~\cite{Cohen-Tannoudji77}: For atom 1, $C_j=\widetilde{C}_j^1 \otimes
\openone_2$, and  for atom 2, $C_j=\openone_1\otimes\widetilde{C}_j^2$.

For example,  
if we denote the jump of atom 1 from
$|e_{-1/2}\rangle_1$ to $|g_{-1/2}\rangle_1$ as channel 2,
the jump operator in the product state basis for this jump is
\begin{eqnarray}
     C_2&=&\sqrt{2/3}\sqrt{\Gamma}\left\{ |g_{-1/2} \rangle_1~|g_{-1/2}
     \rangle_2~_1\langle e_{-1/2}|~_2\langle g_{-1/2}| \right. \nonumber \\
     &&+|g_{-1/2}\rangle_1~|g_{+1/2}\rangle_2~_1\langle
     e_{-1/2}|~_2\langle g_{+1/2}| \nonumber \\
     &&+|g_{-1/2}\rangle_1~|e_{-3/2}\rangle_2~_1\langle
     e_{-1/2}|~_2\langle e_{-3/2}| \nonumber \\
     &&+|g_{-1/2}\rangle_1~|e_{-1/2}\rangle_2~_1\langle
     e_{-1/2}|~_2\langle e_{-1/2}|  \nonumber \\
     &&+|g_{-1/2}\rangle_1~|e_{+1/2}\rangle_2~_1\langle
     e_{-1/2}|~_2\langle e_{+1/2}|  \nonumber \\
     && \left.\! +|g_{-1/2}\rangle_1~|e_{+3/2}\rangle_2~_1\langle
     e_{-1/2}|~_2\langle e_{+3/2}|\right\} \label{eq:C2}
\end{eqnarray}
and the corresponding jump probability for channel 2 is
\begin{eqnarray}
     P_2=\frac{2}{3}\delta t \Gamma
     \left\{ \right. &&|\psi^{e_{-3/2}}_{g_{-1/2}}|^2+
     |\psi^{e_{-3/2}}_{g_{+1/2}}|^2+
     |\psi^{e_{-3/2}}_{e_{-3/2}}|^2  \nonumber \\
     + && \left.  |\psi^{e_{-3/2}}_{e_{-1/2}}|^2+
     |\psi^{e_{-3/2}}_{e_{+1/2}}|^2+
     |\psi^{e_{-3/2}}_{e_{+3/2}}|^2)\right\} \label{eq:P1}.
\end{eqnarray}

We neglect here the case where both atoms jump and two photons are detected
simultaneously. The probability for a single atom jump during $\delta
t$ is $\ll 1$
so the joint jump probability is negligible compared to the single atom jump
probability. In principle it would be possible in simulations to take
into account
joint jumps but this unnecessarily complicates the jump procedure.
After applying
the jump operator $C_j$,  the wave function is still in a
superposition state, but
it has collapsed onto product state basis vectors, leaving only one
ground state
level component of the jumped atom populated.

\subsection{Ensemble averaging}

We calculate the results as an ensemble average of single history
time averages in
the steady state time domain~\cite{Molmer96}. This averaging method requires a
smaller number of histories calculated to achieve reasonable error bars than a
simple ensemble averaging at a single steady state time point. In our study,
extra complications arise because the number of collision processes
in the whole
ensemble also comes into play. Atoms do not end up in the same lattice site,
producing collisions, in all the calculated histories. Atomic hopping between
lattice  sites is a stochastic process and only in a fraction of the
total number of
histories, collision processes occur. We need a sufficiently large ensemble
to produce enough collision events to have reliable results. This is
why we have a
much larger ensemble size, $96-128$
members, than e.g.~used in 3D laser cooling
Monte Carlo simulations using the same ensemble averaging
method~\cite{Molmer95}.

\begin{table}
\caption[t3]{\label{tab:Units} Characteristic units. Distances are given in nm,
momenta in $10^{-28}$ kgm/s, time in ns, and energy in $10^{-30}$ J.}
\begin{tabular}{llllll}
     Quantity
     & Characteristic unit
     & Cs
     & Rb
     & Na
     \\
     \hline
     distance & $\lambdabar=1/k_r$                   & 136  & 124  & 94   \\
     momentum & $p_{r}=\hbar k_{r}$                  & 7.77 & 8.49 & 11.25 \\
     time     & $\Gamma^{-1}$                        & 193  & 167  & 99    \\
     energy   & $E_{r}=(\hbar^{2} k_{r}^{2}) / 2m$   & 1.37 & 2.55 & 16.57 \\
\end{tabular}
\end{table}

In order to be able to use this averaging method, we need to
be sure that
time averages of single histories have been calculated in the steady state time
domain. The simulation times used are displayed in
Table~\ref{tab:SimuT}.
We have carefully checked from the time evolution of the kinetic 
energy that the simulation
time
was long enough to reach well into the steady state time domain.

\section{Simulation scheme} \label{sec:Numerics}

In this section we present the charasteristic units used in our
calculations and
discuss the various criteria which set the numerical limits for the
simulations.
The approximations used and the numerical details related to the wave packet
initial conditions and dynamics are also presented.

\subsection{Scaling and discretization of space}

It is useful from a practical point of view to choose suitable units
and scale the
time-dependent Schr\"odinger equation (\ref{eq:Schrodinger})
accordingly. Convenient
physical units and their numerical values for the three appropriate
alkali metal
elements are listed in Table~\ref{tab:Units}. In the discussion
below, we list all
quantities and scale equations  in units of the characteristic
quantities displayed
in Table~\ref{tab:Units} unless explicitly stated otherwise.

\begin{table}
\caption[t4]{\label{tab:SimuT} Simulation times: Total time for collision
simulations $T$, ensemble averaging time $T_{ave}$, time step size $\delta t$
and maximum momentum $|p|_{max}$ given by $\delta t$ for the numerics to remain
reliable. }
\begin{tabular}{llllllll}
     Simulation
     &  $\displaystyle{T (\Gamma^{-1})}$
     &  $\displaystyle{T_{ave}(\Gamma^{-1} s_0^{-1})}$
     &  $\displaystyle{\delta t (\Gamma^{-1})}$
     &  $\displaystyle{|p|_{max} (p_{r})}$
     \\
     \hline
     Cs374   & 1600 &  78-125  & 0.2  & 110 \\
     Cs584   & 1600 &  97-194  & 0.2  & 110 \\
     Cs1621  &  760 & 178-256  & 0.1  & 155 \\
     Rb560   & 1600 & 140-280  & 0.2  &  89 \\
     Na339   &  470 & 99-198  & 0.05  &  90 \\
     Na530   &  470 & 155-311  & 0.05 &  90 \\
\end{tabular}
\end{table}

As the phase factor $\exp(-iE\delta t/\epsilon)$ has to be well defined,
cf.~Eq.~(\ref{eq:Uformal}), we obtain a criterion for the maximum size of the
time step $\delta t$  dictated by the maximum kinetic energy since we should
fulfill the relation $\delta t \ll\epsilon/p^{2}$. Here $\epsilon$ is
the energy of
the linewidth of the transition in recoil units, as previously displayed in
Table~\ref{tab:AtomProperties}. Collisions increase the atomic kinetic energies
which makes the criterion for $\delta t$ numerically more strict for
the two--atom
case,  compared to the one--atom Sisyphus cooling simulation
(cf.~Fig.~\ref{fig:Localization}). We give in Table~\ref{tab:SimuT}
the values of
$\delta t$ for various simulations and the maximum momentum $|p|_{max}$ for the
numerics to remain reliable.  The total simulation times are $125-311$ in units
of $1 / (\Gamma s_0)$. These depend on the properties of the alkali
metal elements
and the laser parameters.

For the numerical simulations, one has to discretize the position and momentum
spaces, and the resolution has to be fine enough to ensure valid results.
We have used 8192 grid points when the length of the entire spatial grid is
$L_z=5\lambda \simeq 31.4 \lambdabar$. This gives the step sizes in position
and momentum spaces of $\delta z \simeq 0.0038$ and $\delta p=0.2$.
The width of
the Gaussian wave packet at the beginning of the simulation is $\Delta z_0
=0.02\lambda \simeq 0.1257 \lambdabar$ giving $\Delta z_0 / \delta z
\simeq 33$ and
in momentum space $\Delta p_0 / \delta p \simeq 20$, ensuring sufficiently
fine resolution.

The inverse space (here momentum space) has reflecting boundary
conditions when using the FFT method.
Thus the size of the momentum space has to be large
enough to avoid the reflection
of high kinetic energy atoms at the edges of the momentum space.
This requires special attention when considering the interaction
simulations where the kinetic energies of the atoms increase due to
the inelastic
collisions.

The momentum space grid has a total size of $L_k=2 \pi / \delta z =
1638$ so that
the atomic momenta may have values $|p| \leq 819$. The depths of the
lattices in our
simulations are such that  atoms localized at a lattice site have
momenta $|p|<50$.
The momenta increase when the atoms wander around in the lattice,
especially due to
the inelastic collisions. The probability of gaining a sufficient
momentum to reach
the edge of our momentum space grid in a single collision event is
now negligible.
On the other hand many consecutive collisions do not shift the
population for large
$p$ and the reflection effect is avoided.  This is due to the fact that the
increasing relative velocity  between the atoms reduce the excitation
probability
and increases in momentum   terminate before the edges of the $p$--space are
reached.

\subsection{Position fixing of one atom} \label{subsec:Fix}

We simulate the behaviour of a 36 level dissipative quantum system with a
position-dependent coupling to the laser field and a
position-dependent coupling
between two atoms. This requires large computational resources. With
the current
computer capacity, it is not possible to simulate the situation where
both atoms
are allowed to move freely. Instead we have to fix one atom spatially
and let only
the other atom move.
This reduces the dimensionality of the problem
to one since
the relative position of the atoms with respect to the laser field is
now fixed.
This also means that an inelastic interaction process will not change
the kinetic
energy for both atoms, but we use the relative kinetic energy as an
estimate for
the kinetic energy change per atom.

In our previous study~\cite{Piilo01} the position of the fixed atom was kept
constant but here we relax this condition. The position $z_f$ of the
fixed atom is
now  selected randomly  in  
the interval $|z_f| < 0.125\lambda$ for each ensemble
member.  This range covers all the interesting physics as an atom
fixed outside the
current range would be rapidly optically pumped to the opposite lattice well
and the situation would correspond to that with the above-mentioned
range of $z_f$.

The change of $z_f$ also moves the Condon point $r_c$ with respect to
the lattice
and now $r_c$ may be located anywhere between the lattice well  and  
peak. This is an
important point for making our model more realistic:  Since the
kinetic energy of
the atom changes when it moves in  the periodic optical potential, the relative
velocity between the atoms and thus the excitation probability to the
attractive
molecular state  at $r_c$ depends on its position in the lattice. When $r_c$ is
located at the peak of the optical potential, the atom has to move up
the potential
hill to reach it. The relative velocity between the two atoms is now
less and the
excitation probability higher than in the case  where the location of
$r_c$ is at
the bottom of the optical potential.

\subsection{Initial wave packet}

At the beginning of the simulation, the wave packet is placed in a
randomly chosen
ground state sublevel. The initial Gaussian packet has a full spatial width of
$0.02\lambda$.  Thus the initial position of the spatially relatively narrow
wave packet in the lattice is random but completely out of the range of  the
molecular resonance.

Each wave packet has a randomly selected mean initial velocity given by the
Maxwell-Boltzmann distribution corresponding to the selected initial
temperature of
the atomic cloud. We emphasize that the momentum space width of the
initial wave
packet has no association with the thermal distribution, as it is
merely needed to
satisfy the Heisenberg uncertainty relation for a spatially localized initial
state. As stated above, the connection between the wave packet and
the temperature
takes place via the mean momentum of the wave packet. By selecting this mean
momentum randomly for each ensemble member but weighting the
occurrences with the
Maxwell-Boltzmann distribution, we create within the Monte Carlo
ensemble another
ensemble of possible initial collision velocities.
This is the wave
packet version
of the standard collisional energy average~\cite{Machholm01}.

As mentioned above only the ensemble averaged momentum
probability distribution
has a relation to temperature. This   initial
distribution gets narrower when the system evolves and the 
simulation progresses 
corresponding to cooling of the whole atomic cloud.
Moreover, it must be stressed that
the steady state reached does not depend
on the initial widths of the single wave packets
nor on the initial temperature
as long as the atoms are in the reach of Sisyphus cooling.
The simulation times get longer when the initial temperature is increased
but we want to take into account
the effect of collisions on the cooling dynamics in a
realistic way.
We also note that the steady state after cooling in  the  lattice
does not necessarily correspond
to  a  Maxwell--Boltzmann distribution
of velocities but a clear steady state
corresponding to  the 
lattice properties is still reached \cite{Castin90}.

It should be noted that although we include by default the recoil effects by
absorption and stimulated emission, we  cannot in our quantum approach
take proper account of  the Doppler shift which is the basis of the
semiclassical description of Doppler cooling~\cite{Stenholm86}. So in practice
we neglect the Doppler cooling mechanism. Adding the recoil from spontaneous
emission will not change this fact, nor the simulation results. 
In any case,
the role of Doppler cooling is negligible when compared to the Sisyphus cooling
force for the velocities of atoms  localized in the lattice~\cite{Dalibard89}.
It can, however, have a role in the recapture of the hotter atoms heated by
collisions, so in that sense our model is limited.

\subsection{Occupation density of lattice} \label{subsec:Occupato}

We have performed all the simulations presented in this paper for an occupation
density of $\rho_{o}=25\%$ in one dimension, whereas in our earlier
work~\cite{Piilo01}, we presented  also results for other
one-dimensional densities in Cs lattices. These previous simulations 
showed that
an occupation density of 
25\% is sufficiently high for interesting effects to appear, namely an 
evaporative
cooling process which works for at least some parameters of the laser field.
The occupation density used in this paper is also nearly the largest density we
can use when the simulations are done in the way presented here. The purpose of
the paper is to  further explore the parameter space and to extend our
simulations to other atomic masses in addition to describing the details of our
simulation approach.

For $\rho_o=25$\% the available spatial length for the moving atom
should be equal to $\lambda$, corresponding to the average distance between the
atoms. But  decreasing the spatial size increases the step size in the momentum
space since $\delta p = 2\pi / L_z$. So to have a sufficiently fine resolution
in momentum  space and still keep $\rho_o=25\%$, we choose $L_z=5\lambda$ and
set an elastic repulsive potential barrier such that the allowed spatial length
is $\lambda$.  The forbidden spatial region thus makes the numerics work
properly without altering  the physics. Of course the forbidden region does not
affect the momentum space.

Every fourth lattice is occupied when $\rho_o=25$\% or in other
words there is 1
atom per wavelength $\lambda$. This
corresponds to a
situation where the fixed atom sits e.g.\ at $z=0$ and the repulsive elastic
potential barrier for the moving atom is set at $z=\lambda$. Then two
consecutive
collisions are described when the moving atom travels from the first collision
region to the repulsive barrier, turns back, and
collides again.  Memory effects from previous collisions are rapidly
removed due to
decoherence, as also discussed in Sec.~\ref{sec:BinaryI}. This is why
we can say
that the present model describes collisions in general in a lattice
and not only
between the same two atoms.

The dimensionality of the problem and the position
fixing of one atom causes  subtleties  related to the occupation density:
The moving atom travels on average 
 a distance of  $1\lambda$ for the first collision.
After this event it has to travel  a  distance $2\lambda$ (from $z_f$ to 
$z_f+1$ and back) to collide again.
The probability is high for  a  large kinetic energy
increase during the first collision. Thus the first collision
has the dominant effect on  the  kinetic energy scale relevant to 
 the  lattice
dynamics. This is why we rather use here $\rho_o=25\%$ which corresponds
to the average collision distance of $1\lambda$.

It is not trivial to connect the
one-dimensional occupation density to the two- or three-dimensional density, as
that will depend on the particular  form 
of the lattice. It suffices to note,
though, that a certain occupation density in one-dimension will
typically correspond to a lower value in higher dimensions, so that e.g. in
three dimensions we expect to see the effects studied here for
three-dimensional occupation densities less than 25\%.

\subsection{Interaction at short range}

The DDI , Eq.~(\ref{eq:VDipAxis}), becomes singular  at short range. The
singularity in $H_V$ is simply removed by replacing $r$ with $r+r_{\rm off}$
and choosing $r_{\rm off}=10^{-8}$. When constructing the time
evolution operator
$U$, we diagonalize $H_V$ and the DDI part in $H_V$ produces the eigenvalue
manifolds corresponding to the attractive and repulsive molecular potentials.

\begin{figure}[tb]
\centering
\psfig{figure=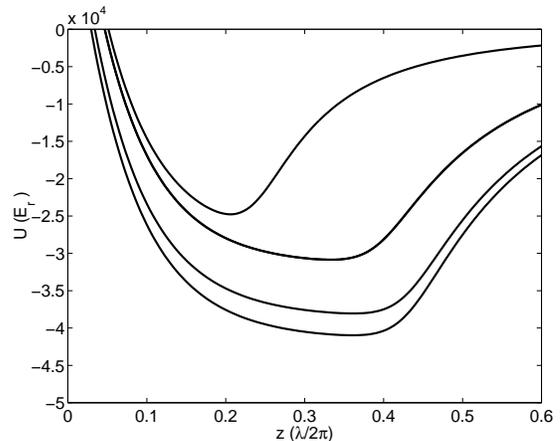,scale=0.4}
\caption[f6]{\label{fig:MolPotsShort}
The short range attractive excited molecular potentials for Cs. Repulsion of
exponential form has been added and the deep part of the potentials 
flattened to
allow reliable numerical treatment of momentum (see text).}
\end{figure}

We replace the position dependencies of the attractive states which
are the same as
in $V_{dip}^{axis}$, Eq.~(\ref{eq:VDipAxis}), by
\begin{equation}
     \frac{1}{r^n} \rightarrow \frac{1}{(r^b+r_{\rm off})^{n/b}},\,
     (n=1,2,3)\label{eq:Vddmani}
\end{equation}
in a manner similar to what was done in Ref.~\cite{Holland94}.
Table~\ref{tab:b}
gives the used values of $b$, and we show the potentials in
Fig.~\ref{fig:MolPotsShort}.

The main reason for this "flattening" of the attractive state potentials is
that considering the numerics, we have an upper limit to momentum.
Thus we need to
set a maximum momentum which can be reached in our simulations by
acceleration, but
which can still be treated reliably numerically in our integration grid, and is
nevertheless large enough to correspond to a clear loss process.  By selecting
different values for $b$ for each molecular potential we take into account the
individual characteristics  of the different attractive states and of
the atomic
elements. It should be noted that by the time the atoms reach the artificially
modified part of the attractive potentials, they move fast enough to make decay
unlikely before they are reflected and move again to the region
where the modification does not affect the potentials 
Thus the flattening of the 
potentials does not
 increase the time 
the atoms spend inside the modified part of the potential, i.e., this 
does not enhance
radiative decay artificially.

We concentrate on the radiative heating which occurs because of the
strong decay in
the vicinity of $r_c$. In this region the treatment of the
singularity in the DDI
and in the attractive molecular states does not yet affect the potentials. For
example, the removal of the singularity causes a change in the value of the Cs
$0_u^+$ potential of 0.7\%  at the position $r=0.50$ when $r_c\simeq 0.83$ for
the detuning used, $\delta=-3\Gamma$.

\begin{table}
\caption[t5]{\label{tab:b}
Values of $b$. Numerical values which are used for the various
attractive molecular
states and atomic species. See also Eq.~(\ref{eq:Vddmani}) and
Figs.~\ref{fig:MolPots} and~\ref{fig:MolPotsShort}. }
\begin{tabular}{llllll}
     element & $0^+_u$ & $1_g$ & $0_g^-$ & $2_u$ & $1_u$
     \\
     \hline
     Cs and Rb     & 22   & 22   & 20   & 20   & 13 \\
     Na            & 13.5 & 13.5 & 12.5 & 12.5 & 10 \\
\end{tabular}
\end{table}

The atoms repel each other at the very short range when their
electron clouds begin
to overlap. We do not have to consider the details of the short range
repulsion.
Thus the short range repulsion is simply produced by adding the
exponential term
$\alpha\,\exp(-\beta r)\,\epsilon$  to the eigenvalues of $H_V$. For
flat states
(states other than attractive or repulsive excited states) $\alpha =
30$, $\beta =
20$ for Cs and Rb, $\alpha = 100$, $\beta = 20$ for Na~\cite{alpha}.  For the
attractive excited state eigenvalue manifolds $\alpha = 25$, $\beta =
15$ for Cs and
Rb, $\alpha = 90$, $\beta = 15$ for Na. Values of $\alpha$ and $\beta
$ are chosen
such that they produce high enough repulsion within a sufficiently
short range but
without producing numerical  difficulties because of the
contradictory requirements
of height and range.

Finally we emphasize that apart from flattening the potentials, we perform our
calculations in the atom-atom basis. Thus the molecular states are not used
directly. Unfortunately our dipole-dipole potential takes care only of the
force part of the atomic interaction. The $r$-dependence of the
lifetimes of the
molecular states are ignored, i.e., each molecular state ends up having the
constant atomic linewidth, instead of the retarded linewidth. This
linewidth arises
from the fact that the two atoms couple to the same vacuum modes at different
locations, leading to an $e^{-i\vec{k}\cdot\vec{r}}$ phase difference
term. However,
the atomic lifetimes differ at maximum only by a factor of 2 from the
atomic one.
The main exception is the 2$_u$ state, which becomes strongly
dipole-forbidden and
can support strong survival and thus e.g. favor the fine structure change loss
mechanism over the radiative process. But we have typically $r_c\lesssim
0.8\lambdabar$, which means that the 2$_u$ state is already hard to excite as
well (see Ref.~\cite{Machholm01} for a detailed discussion).  In
order to make the
quantum jump process tractable we use the atom-atom basis, and,
unfortunately, we
can not just transform into the molecular basis, change the
linewidths into retarded
ones, and then transfer into the atom-atom basis (as we do when flattening the
potentials). This is because for decay, the lifetimes appear in the
jump operators
in addition to the Hamiltonian.

\subsection{Atoms escaping the lattice} \label{subsec:escape}

One needs to define the critical momentum $p_{c}$ to be able to
calculate the MC
ensemble averages and  the properties of the atoms remaining in the lattice.
If $|p|<p_c$ atoms are considered to remain/relocalize in the
lattice, whereas when
$|p|\geq p_c$ they are considered lost from the lattice due to a collision or a
series of collisions.  Semiclassically, we can calculate the critical
$p^{sc}_{c}$ where the cooling force has its maximum value for the parameters
used~\cite{Dalibard89}. The cooling force is, of course, still
effective for momenta
above $p^{sc}_{c}$.

Due to the stochastic nature of the jumps, it is not possible to say
if a given high
momentum atom will relocalize or if it is lost from the lattice. Assume that
an atom has a momentum $|p|>p^{sc}_{c}$. If the following few quantum
jumps reduce
the kinetic energy  of the atom, depending on the atomic position in
the lattice,
it has a good chance to relocalize in the lattice. In the opposite
case, where the
next few jumps increase the kinetic energy of the atom, corresponding
to jumps from
the vicinity of  the bottom of the potential well to the vicinity of
the top of the
well, the atom has less probability to relocalize into the lattice.
This means that
for two different MC histories with the same initial value of $|p|>p^{sc}_{c}$
one atom may escape from the lattice whereas the other one may relocalize.
Thus it is not possible to define $p_c$ in a way that all atoms below
$p_c$ always
relocalize while when $|p|\geq p_c$ they escape.

\begin{figure}[tb]
\centering
\psfig{figure=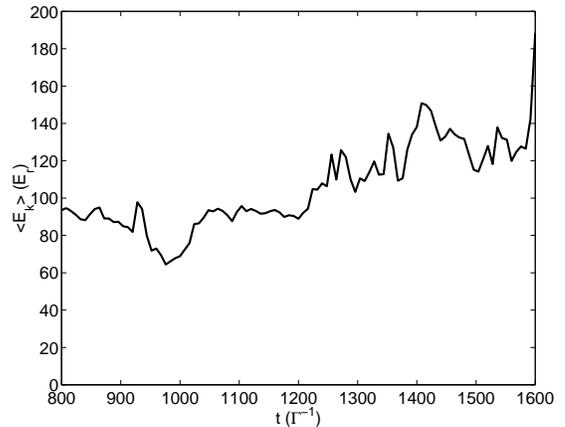,scale=0.4}
\caption[f7]{\label{fig:EkinHighkc}
Kinetic energy time evolution indicating a too large choice for 
$p_c$. The steady
state is not reached since collisions increase the kinetic energy and 
the collided
atoms are out of the recapture range, and thus escape from the lattice. (Cs584,
$p_c=60$).}
\end{figure}

\begin{figure}[tb]
\centering
\psfig{figure=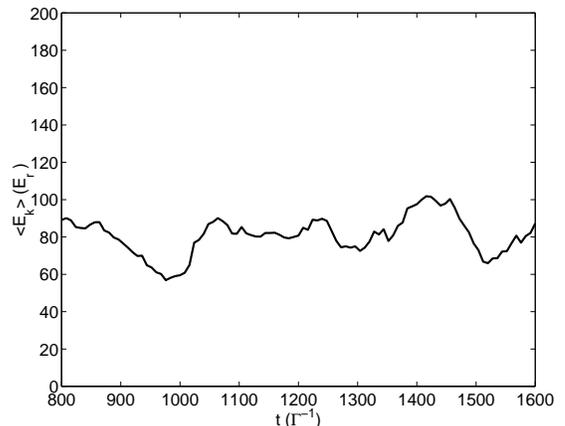,scale=0.4}
\caption[f8]{\label{fig:EkinLowkc}
Kinetic energy time evolution indicating a steady state. Atoms above 
the recapture
range  are now omitted with the proper choice for $p_c$. (Cs584, $p_c=40$). }
\end{figure}

When we calculate the kinetic energy per atom staying in the lattice, we need a
criterion for neglecting those MC histories in the ensemble averaging that
correspond to atoms lost from the lattice. To solve the problem, we
have calculated
the kinetic energy per atom by using various values of $p_c$. Since there is
an increase in the average kinetic energy as a function of time when
the value of
$p_c$ used is too large,  we may check from the time evolution of the kinetic
energy  that our choice for $p_c$ is the proper one when we want to
calculate the
average kinetic energy per atom in the lattice. This is because more collisions
occur as the system evolves in time and if the
gain in kinetic
energy is too large for the atoms to relocalize in the lattice, the
kinetic energy
increases as a function of time and no steady state is reached, as
demonstrated in
Fig.~\ref{fig:EkinHighkc}. Whereas when we use an appropriate value for $p_c$,
atoms still relocalize in the lattice and the kinetic energy exhibits a steady
state behaviour, cf.~Fig.~\ref{fig:EkinLowkc}. It is at the
transition point between
these two different types of behaviours of the kinetic energy that we should
choose the correct value for $p_c$.

Consequently, the atoms of a kinetic energy exceeding the limit given
by $p_c$ are
neglected when we perform the ensemble averaging to find the result
for the kinetic
energy per atom remaining in the lattice. It is important to note that the main
result related to  the narrowing of the momentum probability
distribution due to
collisions still includes all the calculated histories and is totally
independent
of $p_c$.

If there is no constant injection of atoms into the lattice, collisions slowly
deplete it. Finally the density is sufficiently low that the interactions
between atoms are negligible and the atomic cloud regains the
properties determined
by the laser parameters only. It should thus be realized that what we
describe here
is a temporary cooling process which is not effective when the density has
decreased. What we emphasize is the unexpected behaviour of the system in the
intermediate regime where the effect of collisions is not heating but
cooling. This
does not represent the nature of the complete dynamics of the atomic cloud, of
course, as there are other mechanisms, such as the radiation pressure
from scattered
photons, for both heating and cooling, which are not included here.

\subsection{Computational resources} \label{subsec:Comp}

The numerical simulations are demanding since we are dealing with a
36 level quantum
system including various position dependent couplings and dissipative
coupling to
the environment. We use 32 processors of an SGI Origin 2000 machine
which has  128 MIPS
R12000 processors of 1 GB memory per processor~\cite{CSC}. The total
memory taken by a
single simulation (fixed $\delta$, $\Omega$, $\rho_o$ and atomic
species) is 14 GB
and generating a single history requires 6 hours of CPU time. A 
simulation of 128
ensemble members then requires a total CPU time which is roughly 
equal to one month. The
normal clock time is, of course, much shorter  (roughly 22 hours)
since we take
advantage of powerful parallel  processing.

\section{The Semiclassical Approach}\label{sec:SC}

In this section we describe the semiclassical approach to calculate the
excitation and survival probabilities on the  molecular excited states of our
two-atom system.

\subsection{Landau--Zener formula and classical path approximation}

One can calculate the semiclassical excitation probability of the wave packet
traveling through the crossing region between the two states of the
system by using
the Landau--Zener formula
\begin{equation}
     P_{LZ}=1-\exp(-\pi\Lambda) \label{eq:Plz}
\end{equation}
where $\Lambda$ includes both the coupling between the two states and the $C_3$
factor which gives the inverse cubic $r$-dependence of the excited
state ($C_3 /
r^3$)~\cite{Weiner99}.  The basic idea here is that the short resonance
region is approximated to consist of two spatially linearly behaving states
and each component of the wave packet arriving at the resonance region is
independently excited. A more detailed description can be found in
Ref.~\cite{Weiner99}.

One can then calculate the time it takes to reach a point $r$ on the
excited state
by using the classical path approximation
\begin{equation}
     t=t(r)=-\int_{r_{c}}^rdr'\left[\frac{2}{m}
     \left(\frac{p_{cr}^2}{2m}+\frac{C_3}{r'^3} -\hbar \delta
     \right)\right]^{-1/2}\label{eq:Tclass}
\end{equation}
where $p_{cr}$ is the momentum at the Condon point $r_{c}$. There is a direct
correspondence between the reached point $r$ on the excited state and
the energy
gain while accelerating on the attractive excited molecular state. By using
Eq.~(\ref{eq:Tclass}) one can calculate classically how long it takes to reach
a point $r$ corresponding to a given increase in kinetic energy or
momentum due to
the acceleration on the attractive excited state.

It is now easy to numerically calculate the kinetic energy increase due to the
collisions if the wave packet stays on the attractive excited state for a time
corresponding to the natural decay time $\Gamma^{-1}$. When the exponential
decay from the excited state is also taken into account, the probabilities for
various kinetic energy gains due to collisions as a function of
relative velocity
of the colliding atoms at $r_c$ may also be calculated~\cite{Holland94}.

\begin{table}
\caption[t6]{
\label{tab:Plz}
Semiclassical 
probabilities to gain kinetic energies on various
attractive excited
state molecular levels, see Fig.~\ref{fig:MolPots}, for 
our $Cs584$ simulation.
The probabilities are calculated for $p_{tot}=p_{cr}+\Delta p=40$ and
$p_{cr}=24$, this value of $p_{cr}$
corresponds to the lattice depth. The wave packet spends a 
time $t\geq t_{e}$
on the excited state, $P_{LZ}$ is the Landau-Zener excitation probability
and $P_{a}$ gives the survival probability. $P_{tot}=P_{a}P_{LZ}$ is
the total semiclassical
probability  for the process $p>p_{tot}=40$ to occur.}
\begin{tabular}{llllllll}
     potential
     & $\displaystyle{t_{e} (\Gamma^{-1})}$
     & $\displaystyle{P_{LZ}}$
     & $\displaystyle{P_{a}}$
     & $\displaystyle{P_{tot}}$
     \\
     \hline
     $1_{u}$             & 0.21 & 0.71 & 0.81 & 0.57 \\
     $2_{u}$,$0^{-}_{g}$ & 0.40 & 0.88 & 0.67 & 0.59 \\
     $1_{g}$             & 0.48 & 0.91 & 0.62 & 0.57 \\
     $0^{+}_{u}$         & 0.49 & 0.92 & 0.61 & 0.56 \\
\end{tabular}
\end{table}

\subsection{Post collision momentum in the lattice}

\begin{figure}[tb]
\centering
\psfig{figure=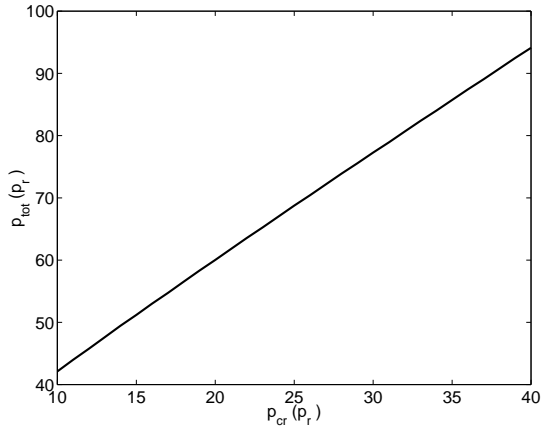,scale=0.4}
\caption[f9]{\label{fig:Ktot}
The total momentum $p_{tot}=p_{cr}+\Delta p$ as a function of resonance point
momentum $p_{cr}$ for the attractive $1_u$ state by the 
semiclassical calculation.
The wave packet has spent a duration corresponding to $\Gamma^{-1}$ 
accelerating on  the 
excited state before spontaneous decay back to the ground state. (Cs584).}
\end{figure}

We obtain the values of $C_3$ for the attractive potentials by fitting near the
resonance region the simple expression $-C_3/r^3$ to  our molecular potentials
obtained by diagonalizing  the dipole--dipole coupling presented in
the two--atom
basis, Eq.~(\ref{eq:VDipAxis}). Figure~\ref{fig:Ktot} shows an
example of the total
post collision momentum $p_{tot}=p_{cr}+\Delta p$ as a function of $p_{cr}$
when the wave packet spent a
time $t=\Gamma^{-1}$ on the $1_u$ excited state ($Cs584$).
One notices that $p_{cr}=24$ corresponding to the
lattice depth used
already gives a total momentum of $p_{tot}=67$ after a collision thus
pushing the
atom to the region in momentum space where  its probability for relocalizing
back to the lattice is small ($p_{c}^{sc}=16.2$). This shows in a 
clear way that
increases in kinetic energy that are large compared to the latttice modulation
depth $U_0$ may occur on a time scale of $\Gamma^{-1}$.

Moreover, when the exponential decay and $P_{LZ}$,
Eq.~(\ref{eq:Plz}), are taken
into account, one is able to calculate the probabilities to gain various
amounts of kinetic energies due to the collision. The total
probability $P_{tot}$
for the atomic momentum to have at least the value $p_{tot}$ after
the collision is
\begin{equation}
     P_{tot} = P_a  P_{LZ}
\end{equation}
where $P_a$ gives the survival probability on the excited state. An
example of the
results of $P_{tot}$ are shown in Table~\ref{tab:Plz}. This suggests
that the first
resonance molecular potential $0^+_u$ has a dominant role in collisions. For
$p_{tot}=40$  the  probabilities for 
 the  various states are roughly equal but
if the first
resonance potential excites and accelerates half of the colliding
atoms only half of
them is left for the remaining potentials.

The simple semi-classical calculation above is not able to give
quantitative results
but it shows that the probability to produce an atom of large momentum due to a
collision is high already when  we consider one excited level only.
This probability
increases when we take into account that during one collision
process, the molecule
may be excited at four different values of $r_c$ related to five different
attractive states.

\section{Simulation Results}\label{sec:Results}

\begin{figure}[tb]
\centering
\psfig{figure=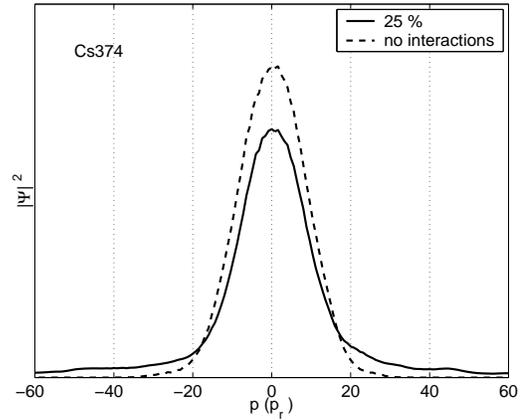,scale=0.4}
\caption[f10]{\label{fig:KCs374}
Momentum probability distributions for interacting and non--interacting cases.
(Cs374).}
\end{figure}

\begin{figure}[tb]
\centering
\psfig{figure=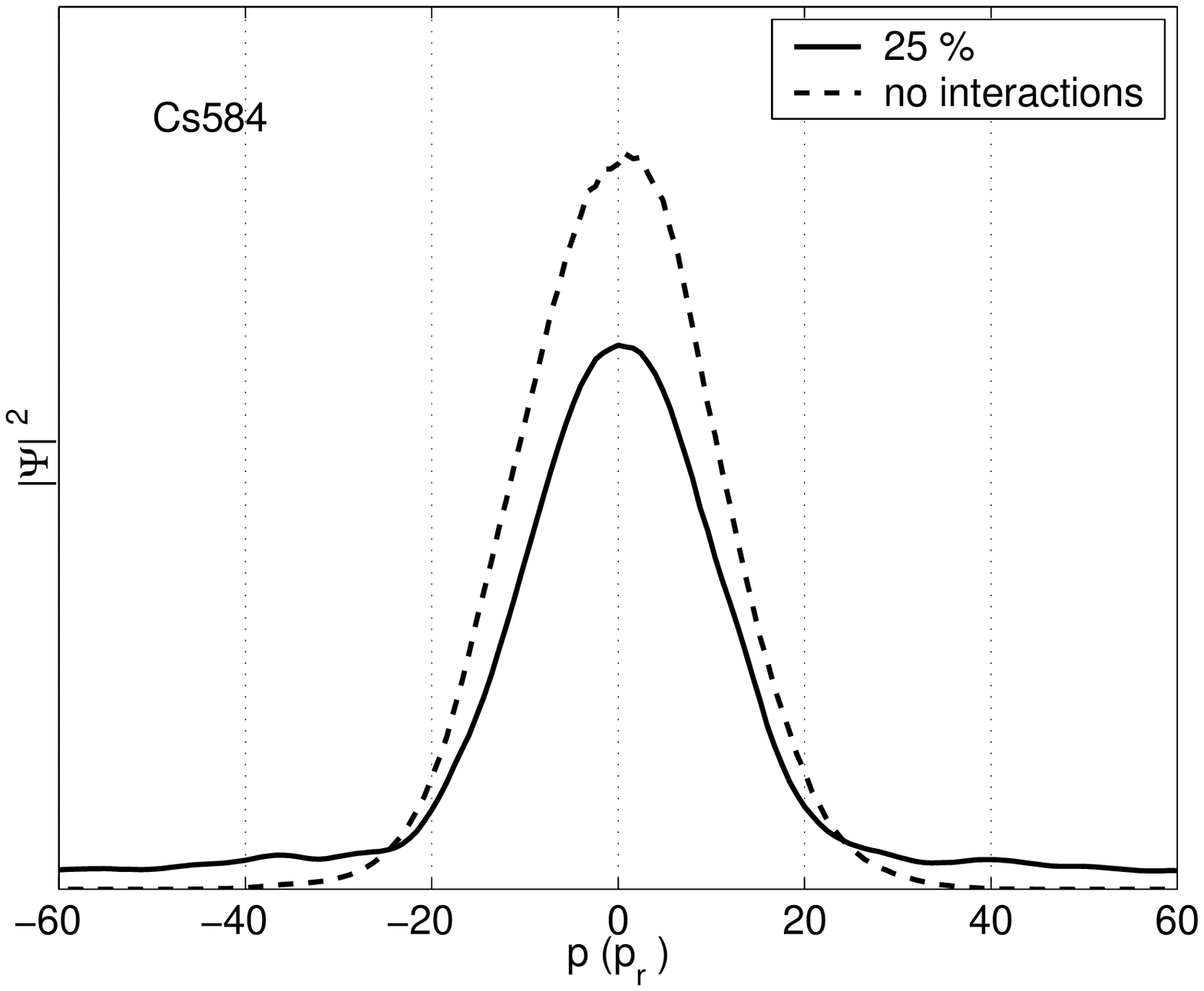,scale=0.4}
\caption[f11]{\label{fig:KCs584}
Momentum probability distributions for interacting and non--interacting cases.
(Cs584).}
\end{figure}

\begin{figure}[tb]
\centering
\psfig{figure=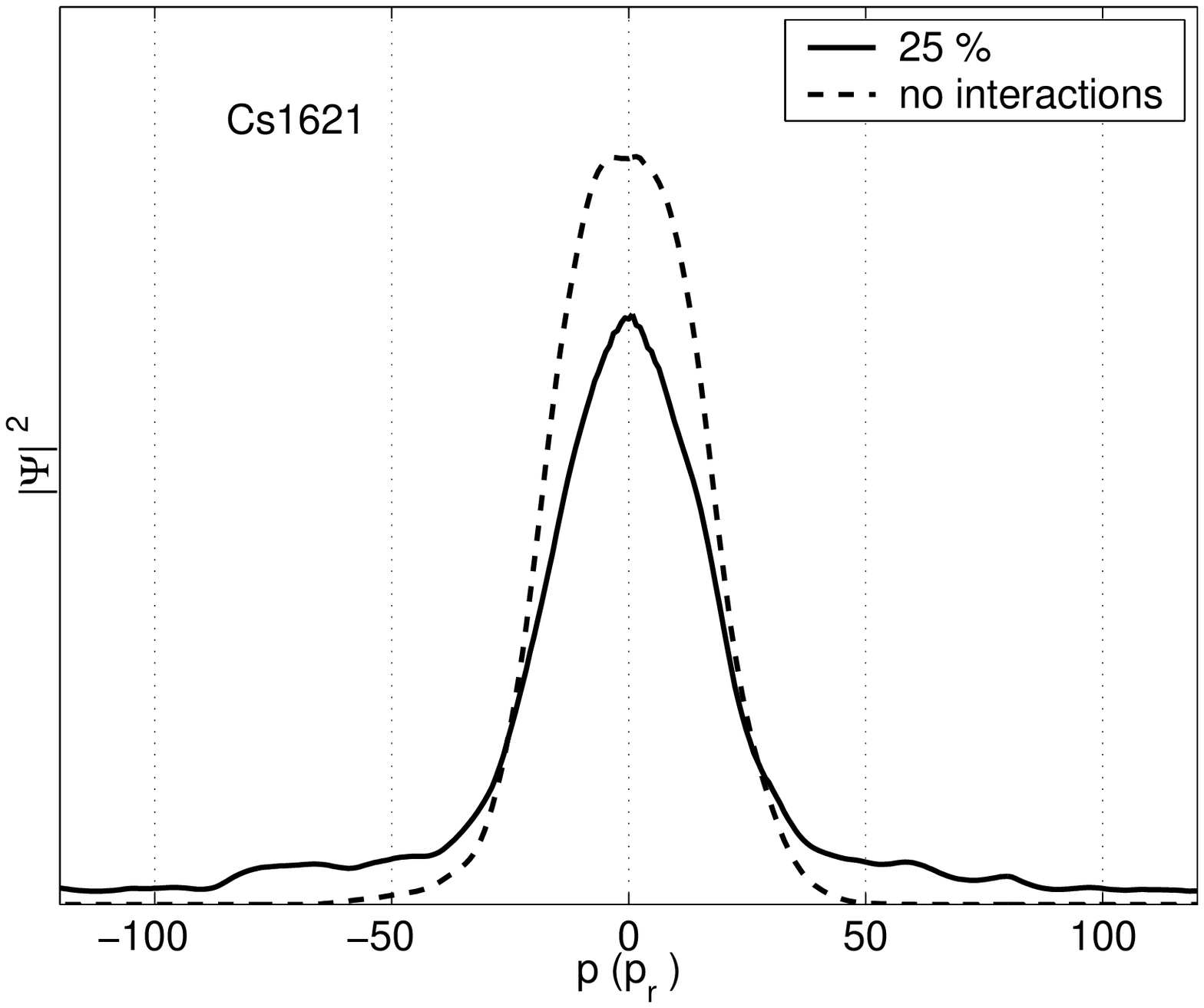,scale=0.4}
\caption[f12]{\label{fig:KCs1621}
Momentum probability distributions for interacting and non--interacting cases.
(Cs1621).}
\end{figure}

\begin{figure}[tb]
\centering
\psfig{figure=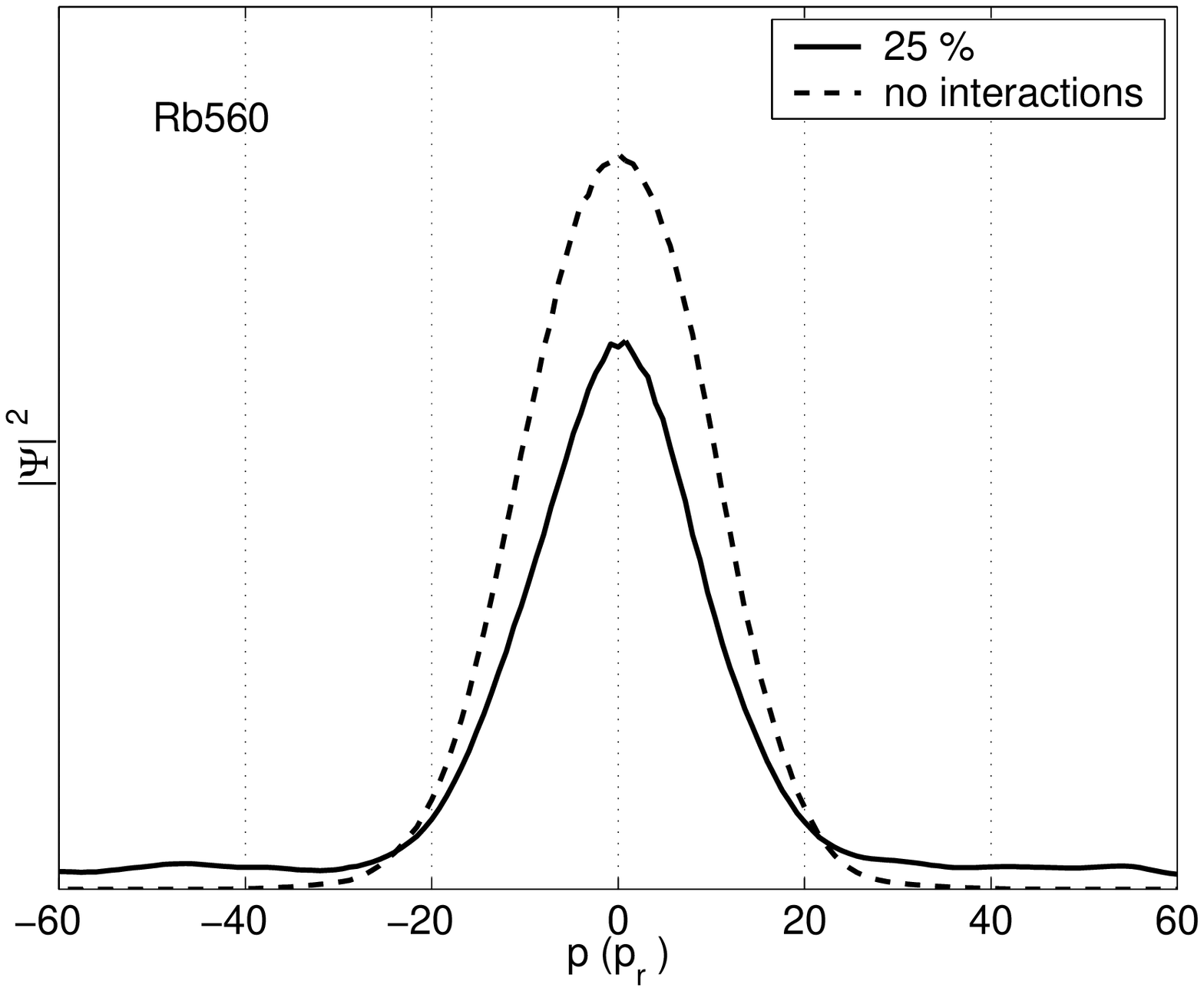,scale=0.4}
\caption[f13]{\label{fig:KRb560}
Momentum probability distributions for interacting and non--interacting cases.
(Rb560).}
\end{figure}

\begin{figure}[tb]
\centering
\psfig{figure=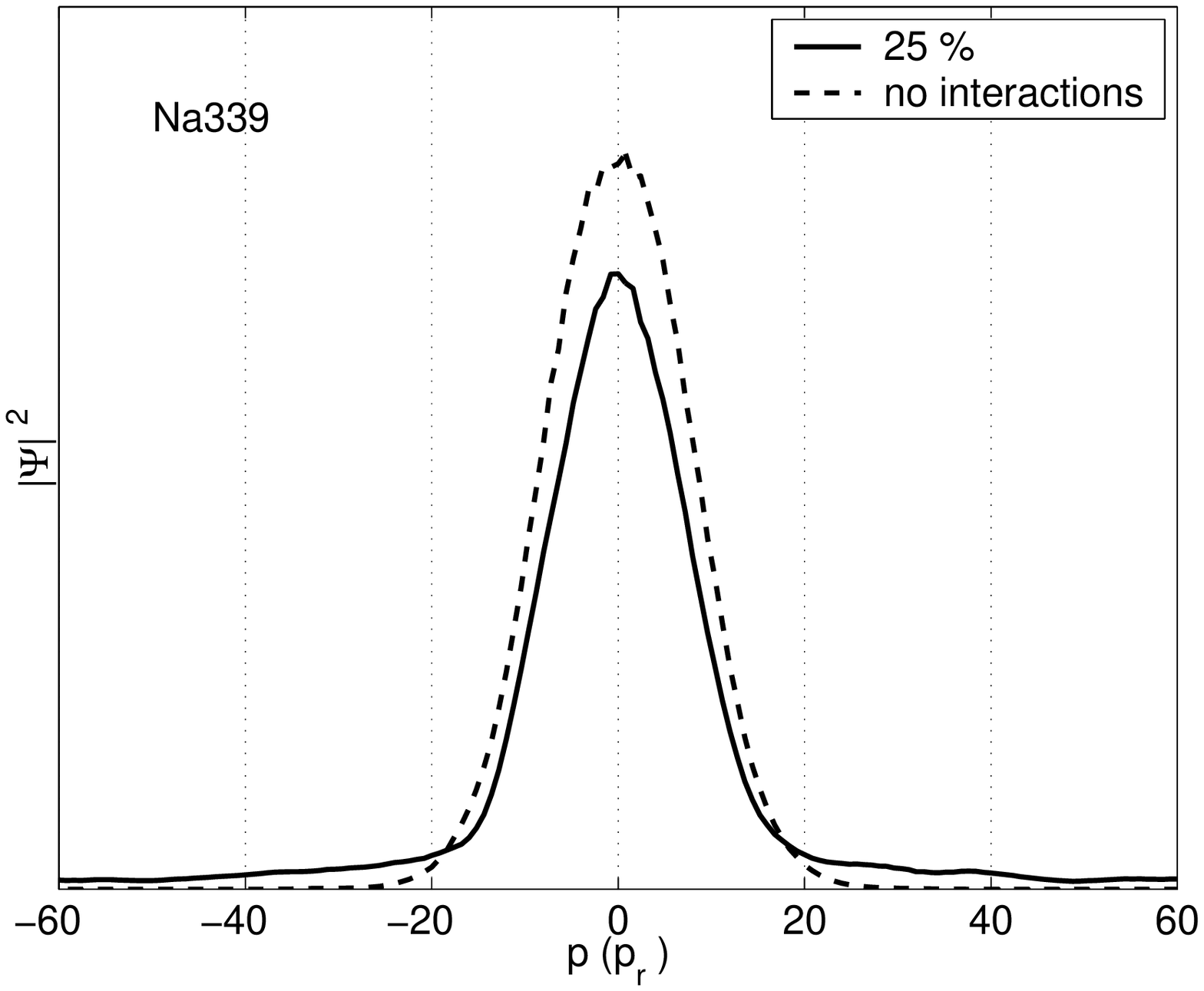,scale=0.4}
\caption[f14]{\label{fig:KNa339}
Momentum probability distributions for interacting and non--interacting cases.
(Na339).}
\end{figure}

\begin{figure}[tb]
\centering
\psfig{figure=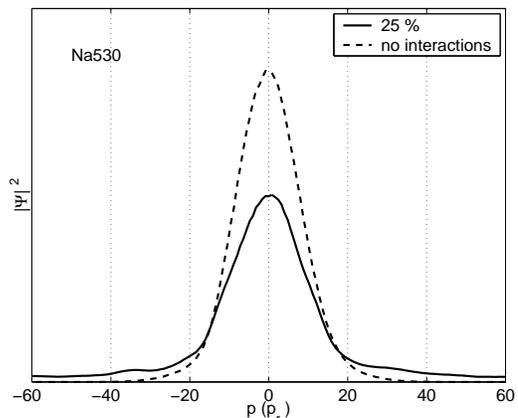,scale=0.4}
\caption[f15]{\label{fig:KNa530}
Momentum probability distributions for interacting and non--interacting cases.
(Na530).}
\end{figure}

The calculated numerical values of kinetic energy per atom for
various simulations
are shown in Table~\ref{tab:Ekin} and corresponding momentum probability
distributions in Figs.~\ref{fig:KCs374}--\ref{fig:KNa530}.

Most of the simulations with the selected parameters produce a reduced value
for the kinetic energy per atom when the interactions between the
atoms are taken
into account,  see Table~\ref{tab:Ekin}.  Since the inelastic
collisions here always
{\it increase} the kinetic energy of the atoms via the radiative 
heating mechanism,
our results suggest  that the consequence of collision almost every time
is the escape
of the colliding high energy atoms from the lattice~\cite{escapeN}.
The atoms left
in the lattice then have less average energy. This is due to the fact
that the more
energetic atoms are favoured to participate in the collision process
due to their
better ability to move between the lattice sites.

The cooling process is indeed observed when looking for the momentum
probability
distributions including all the MC histories for ensemble averaging,
see Figs.~\ref{fig:KCs374}--\ref{fig:KNa530}. One can see the slight
narrowing of
the momentum distributions corresponding to the cooling process. The 
narrow central
peak corresponds to atoms localized in the lattice sites and
the broader background
wing to atoms which are above the recapture range and do not
relocalize in the
lattice. This resembles the evaporative cooling process with narrowed central
peak and  hot background atoms. Cooling here is not dramatic but still
present. Moreover, the result is in sharp contrast when compared to
the theoretical
and experimental collision studies  in MOT's where the heating of the
trapped atoms
due to the radiative mechanism is observed~\cite{Weiner99,Holland94}
but not the evaporative-type cooling process.

The cooling process is observed for all the three atomic masses when similar
lattice depths are used. The computational resources that simulations require,
see Sec.~\ref{subsec:Comp}, do not allow any extensive exploration of parameter
space but the Cs1621 result shows that with a deeper lattice the situation may
change, see Table~\ref{tab:Ekin}. In shallow lattices the relative velocity
before a collision is small, thus enhancing the excitation probability. In deep
lattices the reduced excitation probability due to large relative velocity is
compensated by the use of more intense lasers. The Cs1621 result suggests that
in deeper lattices one may observe heating
which is similar to the
results from MOT studies. But a systematic study of this is out of the reach
of this paper.

\begin{table}
\caption[t7]{\label{tab:Ekin}
Expectation values of kinetic energy per atom ($<E_k>$) for the 
simulations. The
value  of $p_c$ gives the critical momentum which is used in ensemble
averaging to
neglect atoms which have escaped from the lattice.
The absolute values of the standard deviation are given in parentheses.}
\begin{tabular}{lllll}
     Simulation
     & $\displaystyle{p_{c}}$
     & $\displaystyle{<E_{k}> (E_{r})}$
     & $\displaystyle{<E_{k}> (E_{r})}$
     \\
     & & interactions & no interactions
     \\
     \hline
     Cs374  & 35 & 62  (5)      & 75 (5)  \\
     Cs584  & 55 & 82  (6)      & 110 (7)  \\
     Cs1621 & 70 & 264 (30)     & 221 (18) \\
     Rb560  & 50 & 86  (8)      & 95  (5) \\
     Na339  & 40 & 46  (3)      & 59  (3)\\
     Na530  & 45 & 63  (6)      & 84  (6)\\
\end{tabular}
\end{table}

\section{Discussion and Conclusions}\label{sec:Discussion}

Our results show the basic aspects of one mechanism affecting the
thermodynamics of
the atomic cloud in an optical lattice, when the lattice has been prepared with
near-resonant (detuning a few linewidths), red-detuned laser light.
In this case
the role of inelastic collisions is strong, leading to heating and
loss of atoms,
but this requires that the interacting atoms are located in the same
lattice site.
This, on the other hand, requires, firstly, large atomic densities.
Therefore in
most lattice studies done so far, the role of collisions has been
negligible due
to the low densities, or at least not easy to observe (with the exception of
collisions producing a clear signal such as Penning ionisation of colliding
metastable rare gas atoms \cite{Lawall98}).

Secondly, frequent collisions require clear mobility of atoms. This takes place
naturally during the Sisyphus cooling until the atoms are localized in lattice
sites. Thus it is important for suitably dense samples to study the role of
collisions during the Sisyphus cooling, and our approach provides a
method which
is both dynamical and consistent. For the selected parameters our
simulations show
that the Sisyphus cooling process and localization of atoms is not
prohibited by
inelastic processes, i.e., the loss and heating of atoms remains
small even when
the average distance between the two atoms is only four lattice sites.

Once the localization in lattice sites has been achieved as a steady state, the
question about the mobility of atoms changes to some extent. It
should be noted that
localization does not mean that an atom remains in the same site {\it ad
infinitum}. In the steady state the atoms are localized at the sites
for most of
the time, but also move between the sites via tunneling (in the picture where
the lattice lasers and the excited states are eliminated from the effective
description). For the selected parameters the dipole-dipole
interaction does not
perturb the lattice potentials enough to have a significant effect
between atoms
located at different sites (the opposite situation is also possible, see
Ref.~\cite{Boisseau96}). The tunneling of atoms between sites is in the steady
state the main process leading to inelastic collisions, and as the
simulations show
(supported by the semiclassical estimates) such encounters lead
mainly to loss of
hotter atoms or their selective heating. This is because the hotter atoms,
naturally, move between the sites more frequently than the colder
ones, creating
the selectivity. If the lost atoms are not considered, our results
show, however,
that we can hold on to the concept of an existing steady state.

The collisionally induced velocity-selective loss of hot atoms is
similar to the
evaporative cooling which
is utilised in
magnetic traps to reach ultracold temperatures for atoms. It remains
to be seen,
however, whether the resulting cooling effect for the atoms is significant
enough to be observed in densely filled lattices. In dense samples
other processes
such as reabsorption of photons scattered by the atoms are another
important source
for heating, which may well overcome any cooling effect. It should be noted,
however, that if we ignore the spatial structure of the cooling fields, the
collisional processes lose their velocity-selective nature, and as seen in the
simulations of Ref.~\cite{Holland94}, this leads to strong heating of
atoms. Here
the simulations indicate that the lattice structure inhibits this
heating clearly.

Our simulations have been very intense computationally, which makes it very
difficult to make the model more realistic. Our studies, however, to
our opinion,
demonstrate the basic features to be expected from the collisions in densely
populated near-resonant red-detuned lattices. There are
magnetic-field-assisted cooling schemes for blue-detuned lattices for e.g.
$J=1\rightarrow J=1$ systems. The blue detuning usually leads to optical
shielding, and the collisional contribution to inelastic processes is reduced
strongly for normal laser cooling intensities  as the loss channel is
expected to be
adiabatically closed,  see Ref.~\cite{Suominen95}. As a future
prospect it will be
interesting to study the qualitative differences due to the color of
the detuning
in collisions between atoms in near-resonant lattices.

\acknowledgments

J.P. and K.-A.S. acknowledge the Academy of Finland (project 43336), NorFA,
Nordita and European Union IHP CAUAC project for financial support, and the
Finnish Center for Scientific Computing (CSC) for computing resources. J.P.
acknowledges support from the National Graduate School on Modern Optics and
Photonics.


\begin{references}


\bibitem{vanderStraten99} H. J. Metcalf and P. van der Straten,
			{\it Laser Cooling and Trapping}, (Springer, 
Berlin, 1999).

\bibitem{Weiner99}    K.-A. Suominen,
                        J. Phys. B {\bf 29}, 5981 (1996);
                        J. Weiner, V. S. Bagnato, S. Zilio, and P. S. Julienne,
                        Rev. Mod. Phys. {\bf 71}, 1 (1999).

\bibitem{Castin91}    Y. Castin and J. Dalibard,
		      Europhys. Lett. {\bf 14}, 761 (1991).

\bibitem{Dalibard89}  J. Dalibard and C. Cohen-Tannoudji,
		                    J. Opt. Soc. Am. B {\bf 6}, 2023 (1989);
                        P. J. Ungar, D. S. Weiss, E. Riis, and S. Chu,
		                    J. Opt. Soc. Am. B {\bf 6}, 2058 (1989).


\bibitem{Jessen96}     P. S. Jessen and I. H. Deutsch,
                        Adv. At. Mol. Opt. Phys. {\bf 37}, 95 (1996);
                        D. R. Meacher,
                        Contemp. Phys.  {\bf 39}, 329 (1998);
                        S. Rolston, Phys. World {\bf 11} (10), 27 (1998);
                        L. Guidoni and P. Verkerk,
		                     J. Opt. B {\bf 1}, R23 (1999).

\bibitem{Jaksch98}    D. Jaksch, C. Bruder, J. I. Cirac, C. W. Gardiner,
                       and P. Zoller,
		                    Phys. Rev. Lett. {\bf 81}, 3108 (1998);
                        D.-I. Choi and Q. Niu,
		                    Phys. Rev. Lett. {\bf 82}, 2022 (1999);
                       M. T. DePue, C. McCormick, S. L. Winoto, S. Oliver, and
                       D. S. Weis,
		                    Phys. Rev. Lett. {\bf 82}, 2262 (1999);
                       A. J. Kerman, V. Vuleti\'c, C. Chin, and S. Chu,
                       Phys. Rev. Lett. {\bf 84}, 439 (2000).


\bibitem{Holland94}   M. J. Holland, K.-A. Suominen, and K. Burnett,
                        Phys. Rev. Lett. {\bf 72}, 2367 (1994);
                        Phys. Rev. A {\bf 50}, 1513 (1994).

\bibitem{Julienne91}  P. S. Julienne and J. Vigu{\'e},
		                    Phys. Rev. A {\bf 44}, 4464 (1991).

\bibitem{Piilo01}     J. Piilo, K.-A. Suominen, and K. Berg-S{\o}rensen,
                        J. Phys. B {\bf 34}, L231 (2001).

\bibitem{Goldstein96} E. V. Goldstein, P. Pax, and P. Meystre,
		      Phys. Rev. A {\bf 53}, 2604 (1996).

\bibitem{Boisseau96}  C. Boisseau and J. Vigu{\'e},
		      Opt. Commun. {\bf 127}, 251 (1996).

\bibitem{Guzman98}    A. M. Guzm\'an and P. Meystre,
		      Phys. Rev. A {\bf 57}, 1139 (1998).

\bibitem{Menotti99}   C. Menotti and H. Ritsch,
		      Phys. Rev. A {\bf 60}, R2653 (1999);
                        Appl. Phys. B {\bf 69}, 311 (1999).

\bibitem{Dalibard92}  J. Dalibard, Y. Castin, and K. M{\o}lmer,
                        Phys. Rev. Lett. {\bf 68}, 580 (1992);
                        K. M{\o}lmer, Y. Castin, and J. Dalibard,
                        J. Opt. Soc. Am. B {\bf 10}, 524 (1993).

\bibitem{Plenio98}    M. B. Plenio and P. L. Knight,
                        Rev. Mod. Phys. {\bf 70}, 101 (1998)
                        and references therein.

\bibitem{Molmer96}  K. M{\o}lmer and Y. Castin,
                      Quantum Semiclass. Opt. {\bf 8}, 49 (1996).

\bibitem{Suominen98}  K.-A. Suominen, Y. B. Band, I. Tuvi, K. Burnett,
                      and P. S. Julienne,
                       Phys. Rev. A {\bf 57}, 3724 (1998).


\bibitem{Castin90}    Y. Castin, J. Dalibard, and C. Cohen-Tannoudji,
                        in {\it Proceedings of Light Induced Kinetic Effects
                        on Atoms, Ion and Molecules}, edited by L. Moi
                        {\it et al.}, (ETS Editrice, Pisa, 1991).

\bibitem{Cohen-Tannoudji77} C. Cohen--Tannoudji, B. Diu, and F. Lalo\"e,
                              {\it Quantum Mechanics} Vol. I
                              (Wiley--Interscience, Paris, 1977),
                              Chapter II F.


\bibitem{Petsas99}    K. I. Petsas, G. Grynberg, and J.-Y. Courtois,
		                    Eur. Phys. J. D {\bf 6}, 29 (1999) and
                       references therein.

\bibitem{Lenz93}      G. Lenz and P. Meystre,
                       Phys. Rev. A {\bf 48}, 3365 (1993).

\bibitem{Berman97}  P. R. Berman,
		                  Phys. Rev. A {\bf 55}, 4466 (1997).

\bibitem{Castin94}  K. Berg-S{\o}rensen, Y. Castin, K. M{\o}lmer, and
                     J. Dalibard, Europhys. Lett. {\bf 22}, 663 (1993);
                     Y. Castin, K. Berg-S{\o}rensen, J. Dalibard, and K.
                     M{\o}lmer,
                     Phys. Rev. A {\bf 50}, 5092 (1994).

\bibitem{Molmer95}    Y. Castin and K. M{\o}lmer,
                        Phys. Rev. Lett. {\bf 74}, 3772 (1995).

\bibitem{Garraway95}  B. Garraway and K.-A. Suominen,
                        Rep. Prog. Phys. {\bf 58}, 365 (1995)
                        and references therein.

\bibitem{Machholm01} M. Machholm, P. S. Julienne, and K.-A. Suominen,
                      Los Alamos archive, physics/0103059, to appear
                      in Phys. Rev. A (2001).

\bibitem{Stenholm86}    S. Stenholm,
                         Rev. Mod. Phys. {\bf 58}, 699 (1986).

\bibitem{alpha}
These are also the values used for all the states (for Cs, Rb and Na
respectively) to produce the needed reflective boundary conditions by the
repulsive  exponential
potential barrier. When using the Crank-Nicholson method to solve
Eq.~(\ref{eq:Schrodinger}) it is easy to use the reflecting boundary
conditions.
Instead we use the FFT-method since it is computationally faster in the system
studied here.

\bibitem{CSC} See http://www.csc.fi for details.

\bibitem{escapeN}    We have also estimated the number of collision
processes by monitoring the quantum flux over average distance 
$z_{a}$ in single
particle MC simulations. This estimate gives similar results as the number of
high energy atoms produced when interactions between atoms have been included.

\bibitem{Lawall98}    J. Lawall, C. Orzel, and S. L. Rolston,
		      Phys. Rev. Lett. {\bf 80}, 480 (1998).

\bibitem{Suominen95} K.-A. Suominen, M. J. Holland, K. Burnett, and
                      P. S. Julienne,
                      Phys. Rev. A {\bf 51}, 1446 (1995).



\end{references}
\end{document}